\documentclass[aps,prl,superscriptaddress,reprint]{revtex4-2}

\usepackage{amsmath,amssymb,graphicx}

\newcommand{\dd}{\text{d}}
\newcommand{\He}{\text{He}}
\newcommand{\erfc}{\text{erfc}}
\newcommand{\kb}{K_{\rm B}}
\newcommand{\TT}{{\cal T}}

\begin{document}

\title{Discrete Sampling of Extreme Events Modifies Their Statistics}

\author{Lior Zarfaty}
\affiliation{Department of Physics, Institute of Nanotechnology and Advanced Materials, Bar-Ilan University, Ramat-Gan 52900, Israel}

\author{Eli Barkai}
\affiliation{Department of Physics, Institute of Nanotechnology and Advanced Materials, Bar-Ilan University, Ramat-Gan 52900, Israel}

\author{David A. Kessler}
\affiliation{Department of Physics, Bar-Ilan University, Ramat-Gan 52900, Israel}

\begin{abstract}

Extreme value (EV) statistics of correlated systems are widely investigated in many fields, spanning the spectrum from weather forecasting to earthquake prediction. Does the unavoidable discrete sampling of a continuous correlated stochastic process change its EV distribution? We explore this question for correlated random variables modeled via Langevin dynamics for a particle in a potential field. For potentials growing at infinity faster than linearly and for long measurement times, we find that the EV distribution of the discretely sampled process diverges from that of the full continuous dataset and converges to that of independent and identically distributed random variables drawn from the process's equilibrium measure. However, for processes with sublinear potentials, the long-time limit is the EV statistics of the continuously sampled data. We treat processes whose equilibrium measures belong to the three EV attractors: Gumbel, Fr\'echet, and Weibull. Our work shows that the EV statistics can be extremely sensitive to the sampling rate of the data.

\end{abstract}

\begin{titlepage}
\maketitle
\end{titlepage}

{\em Introduction.} Extreme value (EV) statistics is a venerable branch of probability theory, which has drawn much interest over the years \cite{Gumbel1,Leadbetter,Kotz,Coles,Haan}. It finds diverse application not only in physics \cite{Bouchaud,Dean1,Antal,Krapivsky,Comtet,Bertin1,Clusel,Dean2,Sanjib1,Biroli,Dean3,Evans,Fyodorov,Gyorgyi1,Ziff,Furling,Gyorgyi2,Bertin2,Gyorgyi3,Sokolov,Wergen,Fortin,Bar,Oshanin,Godec,Vezzani,Buijsman,Wang,Majumdar,Holl,Claude,Grebenkov,Zarfaty,DeBruyne,Mori}, but in many other fields of science as well \cite{Chen,Burton,Rossi,Sornette,Mikosch,Embrechts,Katz,Orr,Naveau,Gradoni,Castillo,Papal,Naim,Chaves,Cheng,Yan,Lorenz,Schuss}. Predicting when the next EV event will occur and of what magnitude it will be is of practical importance, as the extremes are typically the scenarios we are looking forward to, or alternatively, must watch out for \cite{Wergen,Fortin,Chaves,Sornette,Rossi}. Hence, a thorough understanding of EV statistics is crucial. The EV distribution arising from independent and identically distributed (IID) random variables (RV) has various limiting laws when the sample size approaches infinity \cite{Fisher,Gumbel2,Gnedenko,Hall,Giuliano}, in a similar way to central limit theorems for sums of IID RVs \cite{Kolmogorov}. More precisely, the nature of the tail of the underlying distribution of the IID RVs determines the limiting form of the scaled EV's distribution to be either of Gumbel, Fr\'echet, or Weibull form. However, it is clear that for many natural processes, correlations are vital and omnipresent \cite{Majumdar}, hence the assumption that one is dealing with IID RVs is, in most cases, simply wrong \cite{Dean1,Clusel,Gyorgyi1,Ziff,Wergen,Oshanin,Grebenkov,DeBruyne}.

Typically, one measures an extreme of a time series that represents some quantity, be it for example a temperature \cite{Cheng}, the value of a currency \cite{Lorenz}, or the position of an active biological entity \cite{Schuss}. In principle, the series is continuous, and EV models of such continuously sampled (CS) stochastic paths have attracted considerable attention. However, in reality, for any experimental study the amount of data collected and the sampling rate of the measurement devices are both always finite. Thus, the approach that is relevant to real-world applications is to first discretely sample (DS) the path, and then find the maximum of the sampled sequence of data. Is there a major difference between these two sampling methods?

In this Letter, we answer this question in the context of correlated trajectories of a Brownian particle in a force field, modeled by Langevin dynamics. We start with one of the most well-investigated stochastic processes, the Ornstein-Uhlenbeck (OU) model see (also Refs.~\cite{Godec,Kearney}). It describes the motion of an overdamped particle in a harmonic field or, equivalently, the velocity of a damped Brownian particle. Naively, if the time between sampling events is shorter than the relaxation time, then the former should not be expected to play a major role, and we expect to get the CS EV statistics. But, as we show here, for any finite sampling interval this is wrong.

Our remarkable finding is a qualitative nonsmooth transition from DS to CS in the statistics of extremes, which we present first using the OU model. It exists for any positive sampling interval when the overall measurement time is increased, and is not related to a physical change of the system. It strongly affects the mean and variance of the EV distribution, and thus also the typical fluctuations and large deviations of the EVs \cite{Zarfaty}. Nevertheless, for the OU process both DS and CS give rise to a Gumbel distribution for the EV, in the limit of infinitely long observation time, see below.
\begin{figure}
	\includegraphics[width=1.0\columnwidth]{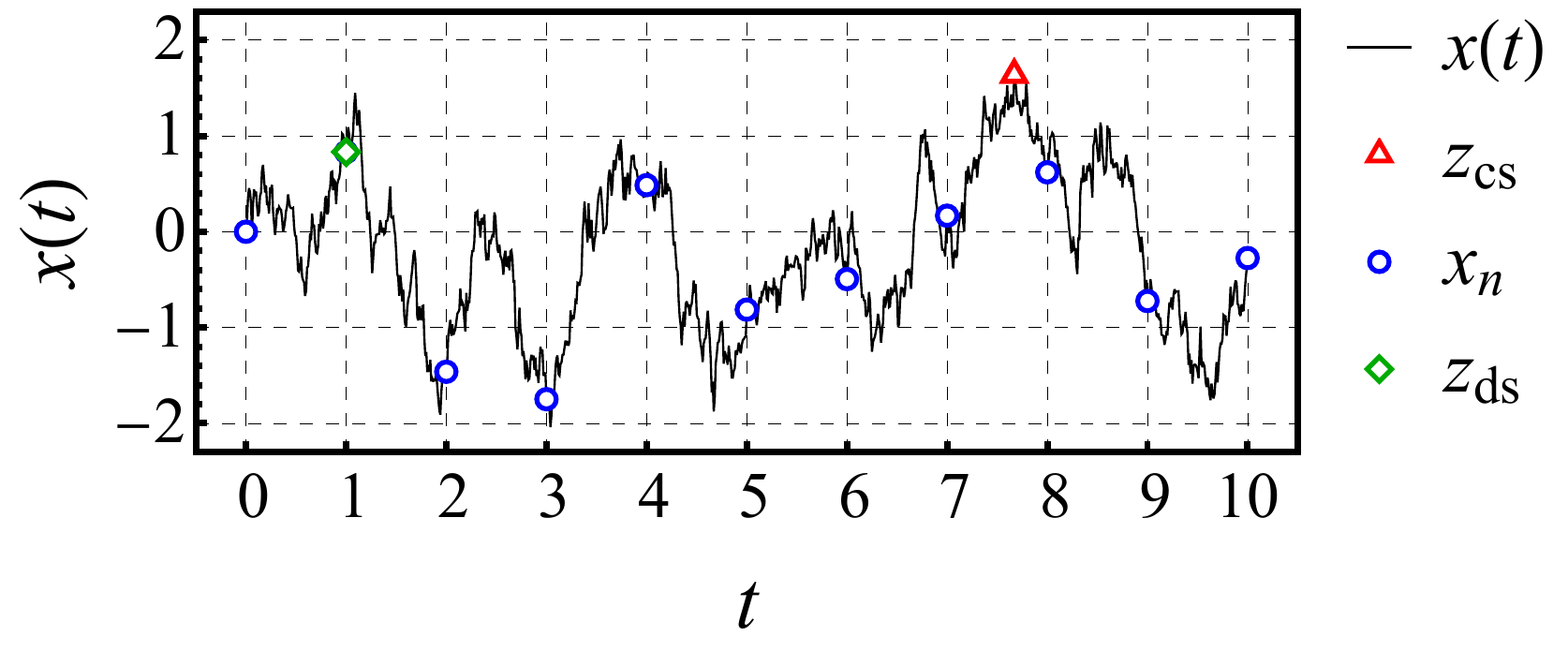}
	\caption{(Color online) {\bf A piece of an OU trajectory:} The path of a Brownian particle in a confining harmonic force field, modeled via the OU process (solid curve). This path's maximum is $z_{\rm cs}$ (red triangle), while discretely sampling every $\Delta=1$ unit of time yields the sequence $x_n$ (blue circles), with a maximum of $z_{\rm ds}$ (green diamond).}
\label{fig1}
\end{figure}

We then extend our results to a wide class of Langevin processes that lie in the Gumbel domain, unveiling a second transition governed by the large-displacement behavior of the force field controlling the dynamics. Finally, within this Langevin approach, we briefly present in the Appendix extensions to processes whose equilibrium distributions (ED) belong to the other two EV limits, Fr\'echet and Weibull.

{\em The OU model.} We start by considering the Langevin equation for the OU model,
\begin{equation}
\label{eq01}
	\frac{\dd}{\dd t} x(t) = - \frac{1}{\tau}x(t) + \sqrt{2D}\eta(t) ,
\end{equation}
where $\tau$, $D$, and $\eta(t)$ are the relaxation time, the diffusion coefficient, and standard Gaussian white noise, respectively. The noise obeys $\langle\eta(t)\eta(t')\rangle = \delta(t-t')$ and has zero mean, where $\delta(\cdot)$ is Dirac's delta function. The particle, at position $x(t)$, is subject to a force which is derived from a quadratic potential. We rescale all quantities in the equation such that $t$ and $x(t)$ are measured in units of $\tau$ and $\sqrt{D\tau}$, respectively. We specialize to this OU path $x(t)$ in the time interval $[0,T]$, and sample it stroboscopically every $\Delta$ units of time; see Fig.~\ref{fig1}. The outcome of this DS measurement is the random sequence $x_n \equiv x(n\Delta)$, where $0\le n\le N$ and $N\Delta=T$ is the total measurement time. We focus on the maximum of this set, denoted $z_{\rm ds}$, and compare its properties to those of the previously studied case of the maximum of $x(t)$ in the interval $[0,T]$, $z_{\rm cs} \equiv \max_{0\le t\le T}[x(t)]$ \cite{Majumdar,Pickands}. To compute this latter quantity, one has to measure the whole continuous trajectory, and hence we call it the CS model. Clearly, $z_{\rm ds} \le z_{\rm cs}$.

The binding force ensures that an ensemble of particles will reach a steady state, the Boltzmann-Gibbs measure, given by $\phi(x)\equiv\exp(-x^2/2)/\sqrt{2\pi}$. In the limit of large $\Delta$ and $T$ but fixed $N$, the sampling is of uncorrelated RVs all drawn from the ED. Thus, if $z_{\rm ds}<z$ then all the $N$ sampled variables are also smaller than $z$, and since they are IID RVs drawn from the ED we find that $\lim_{\Delta\to\infty}\text{Prob}(z_{\rm ds}<z) = [\Phi(z)]^N$, with $\Phi(z) \equiv \int_{-\infty}^z\dd x\,\phi(x) = 1-\erfc(z/\sqrt{2})/2$ and $\erfc(\cdot)$ is the complementary error function. In this limit, the nature of the EV statistics is only due to the equilibrium properties of the system, and any dynamical information, including correlation effects, is wiped out. When $N$ is large, the typical EVs are also large \cite{Zarfaty}; hence we assume $z \gg 1$, where $\Phi(z) \simeq 1-z^{-1}\phi(z)$, and get
\begin{equation}
\label{eq02}
	\lim_{\Delta\to\infty}\text{Prob}\left(z_{\rm ds}<z\right) \sim \exp\left[-Nz^{-1}\phi(z)\right] .
\end{equation}

To treat the DS EV case, we consider the positions $x_n$ at the moments of sampling using a discrete stochastic map. By integrating the Langevin equation, Eq.~(\ref{eq01}), one finds the OU update formula, $x_{n+1} = \mu x_n + \sqrt{1-\mu^2}\eta_n$, where the $\eta_n$s are standard Gaussian IID deviates and $\mu \equiv \exp(-\Delta)$ \cite{Gillespie}. In the large-$N$ limit, we find
\begin{equation}
\label{eq03}
	\text{Prob}\left(z_{\rm ds}<z\right) \sim A(z)\exp\left\{-N\ln\left[\frac{1}{\Lambda_*(z)}\right]\right\} .
\end{equation}
The amplitude $A(z)$ approaches unity for large $z$ and the main focus here is the largest eigenvalue, $\Lambda_*(z)$. The latter obeys the following integral equation, obtained from the stochastic map \cite{Supp},
\begin{equation}
\label{eq04}
	\Lambda_*(z)P_*(x;z) = \int_{-\infty}^z \frac{\dd x'\,P_*(x';z)}{\sqrt{2\pi(1-\mu^2)}} \exp\left[-\frac{(x-\mu x')^2}{2(1-\mu^2)}\right] ,
\end{equation}
where $P_*(x;z)$ is the corresponding eigenfunction. Evaluating the joint limit of $\Delta \to 0$ and $N\to\infty$ with $T$ fixed and large \cite{Supp}, we obtain the Fokker-Planck description of the problem, $\lim_{\Delta\to0}\text{Prob}(z_{\rm ds}<z) \sim \exp[-T\lambda_*(z)]$, i.e., the CS limit, with $\lambda_*(z) \equiv \lim_{\Delta\to0}[1-\Lambda_*(z)]/\Delta$. In Ref.~\cite{Majumdar}, it was shown that $\lambda_*(z)$ is the smallest magnitude solution of $\text{D}_{\lambda_*(z)}(-z)=0$, $\text{D}_{\cdot}(\cdot)$ being the parabolic cylinder function, a result which we recover. For large $z$, one has $\lambda_*(z) \sim z\phi(z)$ \cite{MajumdarPC}, and the CS limit becomes \cite{Majumdar,Pickands}
\begin{equation}
\label{eq05}
	\lim_{\Delta\to0}\text{Prob}\left(z_{\rm ds}<z\right) \sim \exp\left[-Tz\phi(z)\right] .
\end{equation}
The Gaussian decay of the exponents in Eqs.~(\ref{eq02}) and (\ref{eq05}) means that both the IID and CS limits belong to the Gumbel universality class. However, the large-$z$ asymptotic behavior of these two exponents differs by a diverging factor of $z^2$, making the corresponding EV distributions vastly different. Surprisingly, for any finite $\Delta$, the large-$N$ limit of the DS process's EV distribution, Eq.~(\ref{eq03}), which is dominated by the large-$z$ asymptotics of the eigenvalue $\Lambda_*(z)$, converges to the EV measure given by the ED IID limit, both for the OU process along with a wide class of similar processes, as we show below. Hence, the limit of $\Delta \to 0$ is singular in the context of EV theory \cite{Berman}.
\begin{figure}
	\includegraphics[width=1.0\columnwidth]{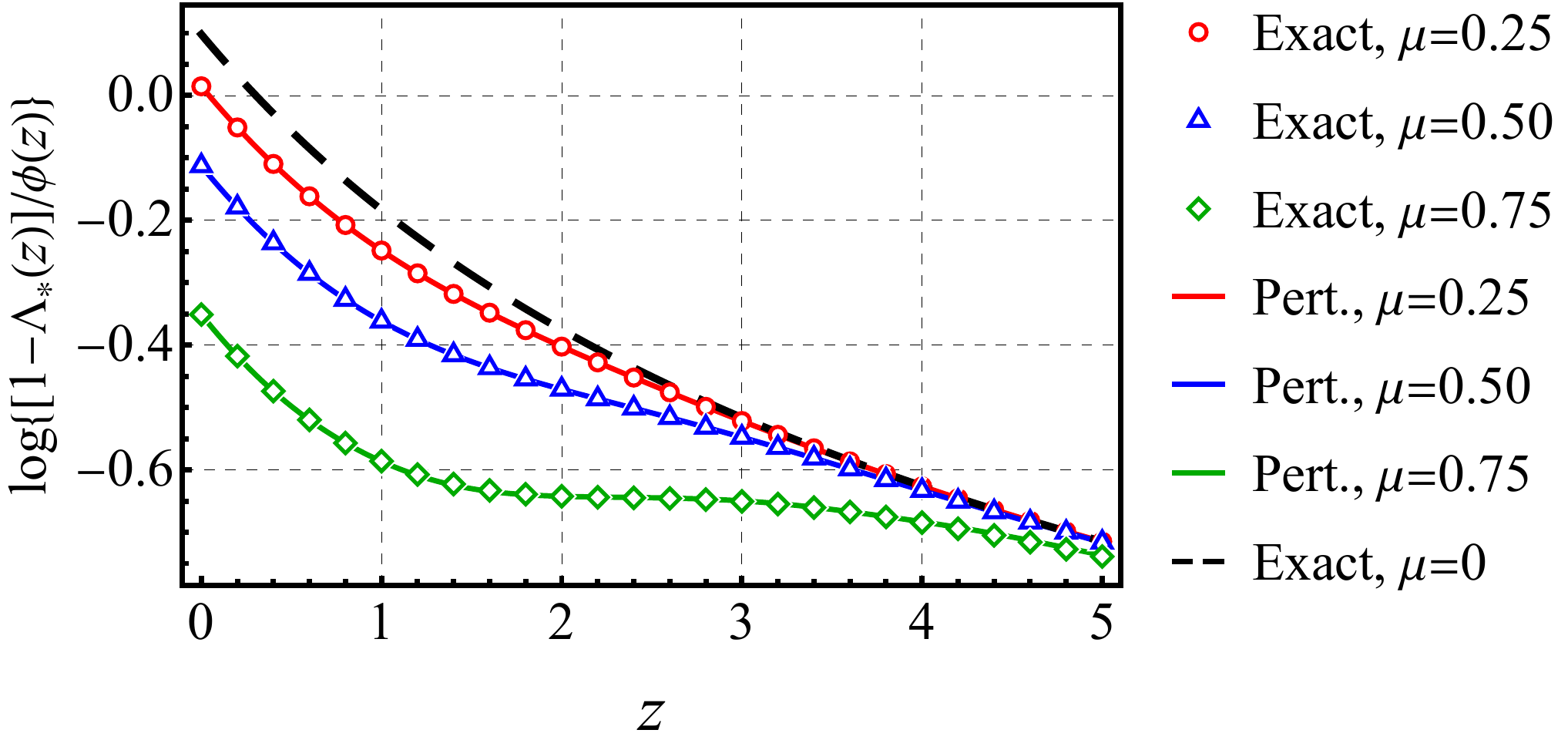}
	\caption{(Color online) {\bf The large-$z$ convergence of $\Lambda_*(z)$:} The scaled eigenvalue $[1-\Lambda_*(z)]/\phi(z)$ for $\mu\equiv\exp(-\Delta)=0.25$ (red circles), $\mu=0.5$ (blue triangles), and $\mu=0.75$ (green diamonds), obtained from numerical evaluations of the eigenvalue equation, Eq.~(\ref{eq04}), as well as from a tenth-order perturbative expansion in $\mu$ (solid curves). Also shown is the exact result for the IID case, $\mu=0$, for which $\Lambda_*(z) = \Phi(z)$ (dashed black line). Notice that all three finite-$\mu$ curves merge for large $z$ with the IID line.}
\label{fig2}
\end{figure}
\begin{figure*}
	\includegraphics[width=1.0\textwidth]{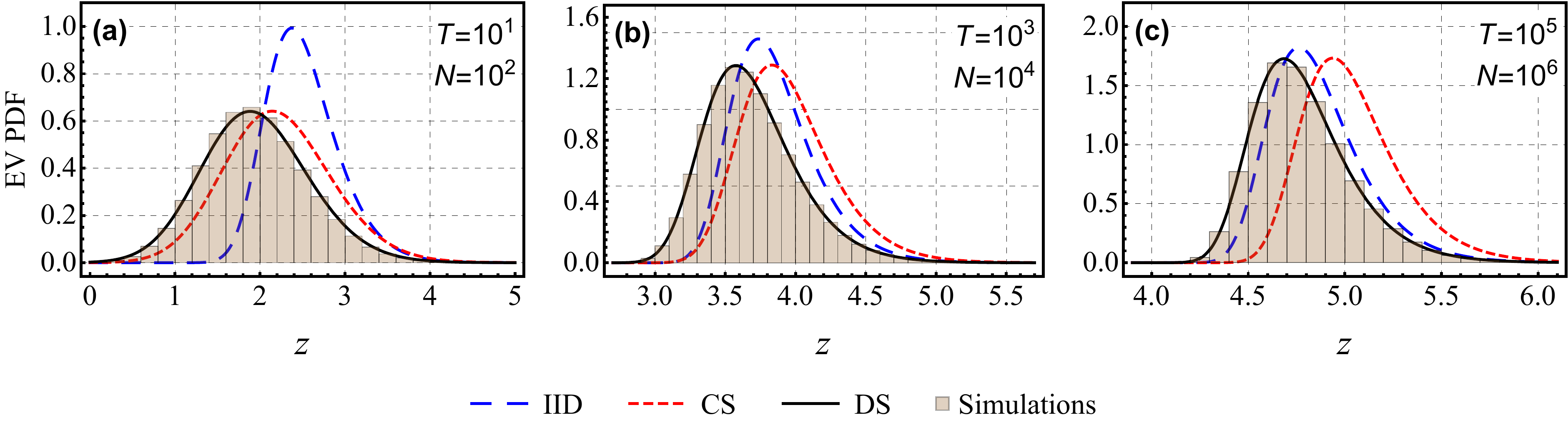}
	\caption{(Color online) {\bf The EV statistics of the DS OU model:} The distribution of EVs for the DS OU process, with sampling rate $\Delta=0.1$. For not too large $T$, we see a behavior close to that of the CS approach (a). However, as we increase $T$, approximating the DS statistics by those of CS becomes less accurate (b), and eventually approach the statistics predicted for $N=T/\Delta$ IID RVs drawn from the ED (c). The IID and DS curves (dashed blue and solid black) correspond to $\exp\{-N\ln[1/\Lambda(z)]\}$ with $\Lambda(z)=\Phi(z)$ and $\Lambda(z)=\Lambda_*(z)$, respectively. The CS curve (short-dashed red) corresponds to $\exp[-T\lambda_*(z)]$, where $\text{D}_{\lambda_*(z)}(-z)=0$. Each histogram is made of $10^6$ maxima with initial conditions of $x=0$.}
\label{fig3}
\end{figure*}

To begin analyzing the DS EV problem, we use a small-$\mu$ (or equivalently, large-$\Delta$) perturbation theory, expanding $\Lambda_*(z) = \sum_{n=0}^{\infty}\lambda_n(z)\mu^n$, and similarly for $P_*(x;z)$. Using Eq.~(\ref{eq04}), we get that $\Lambda_*(z) \simeq \Phi(z)+\mu[\phi(z)]^2/\Phi(z)$ to first order in $\mu$. For large-$z$ this implies that
\begin{equation}
\label{eq06}
	\Lambda_*(z) \simeq 1 - z^{-1}\phi(z) + \mu[\phi(z)]^2 .
\end{equation}
The second term is expected as it is the result obtained for IID RVs that originate from the ED. A key observation is that for large $z$, the third term is by far smaller than the second one, even if $\mu$ is not too small, since $\phi(z) \ll 1$. The first-order correction with $\mu$ is thus exponentially small in $z$ with respect to the leading term. We continue the small-$\mu$ expansion to order $10$ \cite{Mathematica} and find, similarly, that all the terms up to $\mu^{10}$ are negligible in the large-$z$ limit. This behavior is also found in numerical calculations of the eigenvalue $\Lambda_*(z)$ \cite{Supp}, as exhibited in Fig.~\ref{fig2}, showing that for large values of $z$ all the numerical data converge to a unique curve which is $\Delta$ independent, namely the IID curve. This accords with the result of Berman \cite{Berman} for stationary Gaussian sequences, that when $z$ is large the EV statistics will converge to that of IID RVs drawn from the ED for any positive $\Delta$.
\begin{figure}
	\includegraphics[width=1.0\columnwidth]{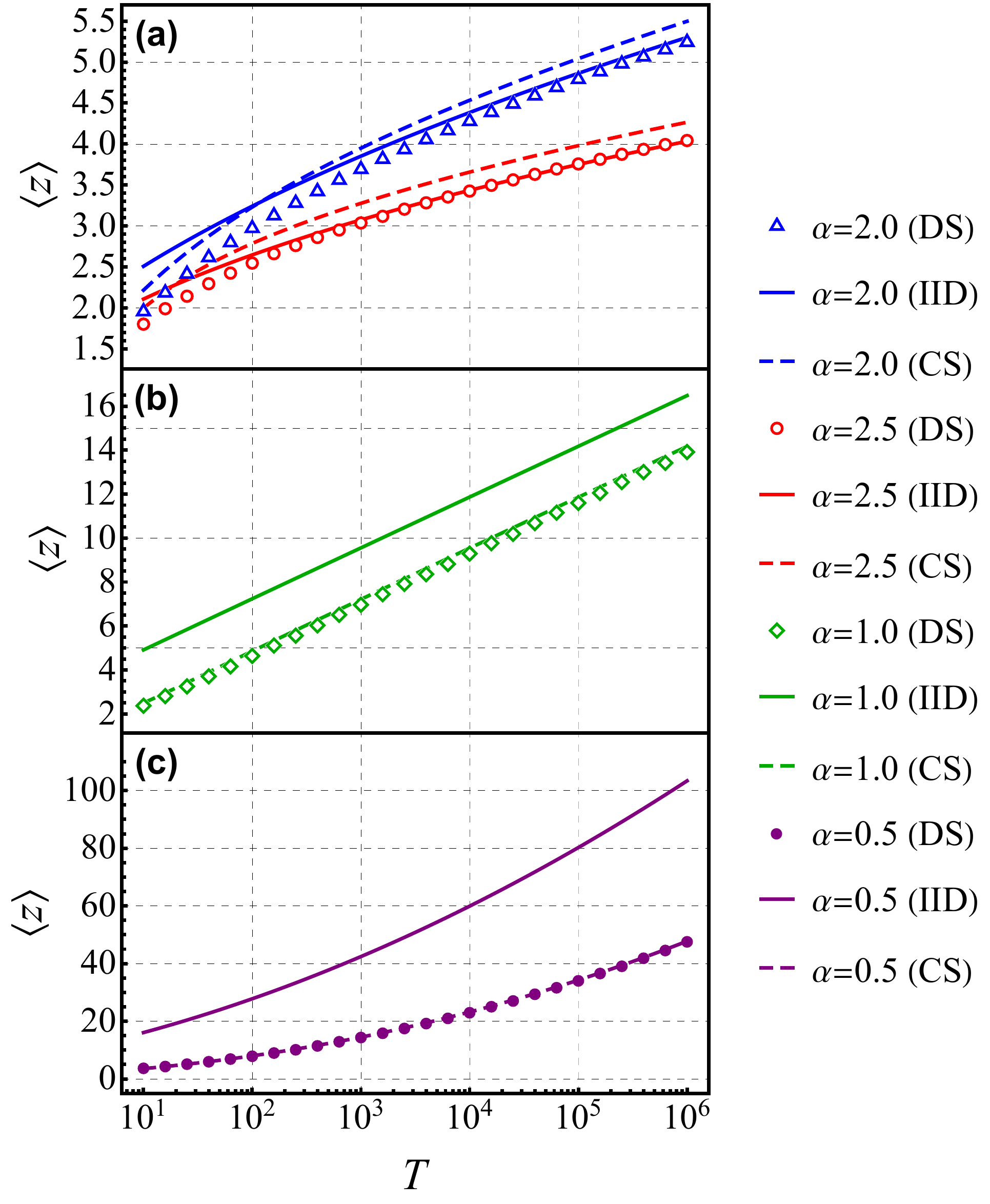}
	\caption{(Color online) {\bf The Gumbel class:} The mean EV $\langle z \rangle$ of a DS process $x(t)$, evolving according to Eq.~(\ref{eq01}) ($D=1$), but with a deterministic force of $-U'(x)$, where $U(x) = (1/\alpha)(1+x^2)^{\alpha/2}$. We used (a) $\alpha=2$ (blue triangles) and $\alpha=2.5$ (red circles), (b) $\alpha=1$ (green diamonds), and (c) $\alpha=0.5$ (purple disks), corresponding to the OU model and to increasing, constant, and decreasing-force processes, respectively. Seen are numerical evaluations for these four cases, where the sampling interval is $\Delta=0.1$. Also depicted are the ED IID (solid lines) and CS (dashed curves) predictions for each value of $\alpha$. (a) For $\alpha>1$, the DS values converge to the IID description. (c) The opposite happens for $\alpha<1$, as this case has a force that vanishes for large distances. (b) The borderline case is $\alpha=1$, where the DS, IID, and CS values do not seem to intersect. Each mean is made of $10^4$ maxima whose initial conditions are $x=0$, obtained using the Euler–Maruyama method with an underlying time increment of $0.01$. A reflective boundary condition at $x=0$ was used when $\alpha<1$.}
\label{fig4}
\end{figure}

To further elucidate this phenomenon, we need a different strategy that exploits the large-$z$ expansion of the integral eigenvalue equation, i.e., Eq.~(\ref{eq04}). Expressing the largest eigenvalue as $\Lambda_*(z) \simeq 1 - \phi(z)\Lambda_1(z) + [\phi(z)]^2\Lambda_2(z)$, and similarly for $P_*(x;z)$, we obtain \cite{Supp}
\begin{align}
\label{eq07}
	\Lambda_*(z) \simeq 1 & -\phi(z) \underbrace{\frac{\erfc(z/\sqrt{2})}{2\phi(z)}}_{\Lambda_1(z)}	\nonumber \\
	& +[\phi(z)]^2 \underbrace{\sum_{n=1}^{\infty} \frac{\mu^n/n!}{1-\mu^n} \He^2_{n-1}(z)}_{\Lambda_2(z)} ,
\end{align}
where $\He_n(\cdot)$ is the $n$th probabilists' Hermite polynomial. Further expanding Eq.~(\ref{eq07}) for large $z$, we find
\begin{align}
\label{eq08}
	\Lambda_*(z) &\simeq 1 - \frac{\phi(z)}{z} + \left[\frac{\phi(z)}{z}\right]^2\frac{(1+\mu)^2}{\sqrt{1-\mu^2}} \exp\left(\frac{z^2\mu}{1+\mu}\right) \nonumber \\
	&\simeq 1 - \frac{\phi(z)}{z} \left(1 - \frac{2e^{-\Delta z^2/4}}{\sqrt{\pi\Delta z^2}}\right) ,
\end{align}
where the last expression is valid for small $\Delta$ \cite{Limit}. Remarkably, the leading two terms are $\mu$ independent and correspond to the result for IID variables originating from the ED. However, for fixed $z$, when $\Delta$ becomes small, or equivalently $\mu$ approaches unity, the last term diverges, indicating the breakdown of the large-$z$ perturbation theory and the existence of a crossover regime to a CS behavior for $\Delta z^2 \sim {\cal O}(1)$. This is evidenced in Fig.~\ref{fig3}, where one sees that for small $T=\Delta N$, the distribution of $z_{\rm ds}$ is close to the CS prediction, whereas for large $T$ it appears to converge to the IID limit. This transition has however nothing to do with a physical switch of the behavior of the system, and is rather a purely statistical effect due to the finite sampling rate. Thus, for any fixed $\Delta>0$, as $T$ becomes large the IID statistics and ED control the EV theory.

{\em A qualitative argument.} How are we to understand the crossover scale of $\Delta z^2 \sim {\cal O}(1)$? A simple explanation to this result is as follows. Let us expand the recursion relation of $x_n$ for small $\Delta$, $x_{n+1}-x_n \simeq -\Delta x_n + \sqrt{2\Delta}\eta_n$. We see that there is a competition between two terms. For small $\Delta$ the stochastic noise is dominant, and so a record-breaking large $x_n$ is very liable to be followed by a yet larger value. However, for sufficiently large $x_n$, the deterministic term which is proportional to $x_n$ dominates, so those maxima are separated by large gaps in time. These two terms are comparable precisely in the crossover regime we have identified. Physically, the effect we find here is related to the fact that extreme events of Langevin paths in a confining field become larger as time progresses. However, the bigger the true maximum is (in the CS sense), the faster the relaxation from this extreme gets, simply because the restoring force field gets enormously large if the path wanders to an EV. This idea suggests that our main result found for the OU process is of more general validity. We explore this by considering the path of a Brownian particle subjected to more general binding force fields. As explained below, these results extend beyond the Gumbel basin of attraction.

{\em Other force fields in the Gumbel domain.} Let us consider a potential of the form $U(x) = (1/\alpha) (1+x^2)^{\alpha/2}$, with $\alpha>0$ (see further details in the Supplemental Material \cite{Supp}). In Fig.~\ref{fig4}, we plot the mean EV $\langle z \rangle$ versus $T$ given various values of $\alpha$. For the OU process with $\alpha=2$, we see that the numerical values converge to the IID limit at large times; see Fig.~\ref{fig4}(a). This works also for $\alpha=2.5$, since here too the force grows with $x$, leading to a domination by the deterministic force term at long times. However, this argument is no longer valid for $\alpha\le1$, where the force does not increase with $x$; see Figs.~\ref{fig4}(b) and \ref{fig4}(c). For example, when setting $\alpha=0.5$, the stochastic term dominates at large $x$, and the exact values (which are nicely described by CS) diverge from the IID behavior; see Fig.~\ref{fig4}(c). When $\alpha=1$, the force is asymptotically constant, which is a special borderline case with all curves being parallel; see Fig.~\ref{fig4}(b). This case was also shown to be critical for problems which do not involve DS; see Ref.~\cite{Sanjib1} in the context of crowding of near-extreme events, and Ref.~\cite{Sanjib2} where a freezing transition was discovered for the long-time decay rates of first-passage probabilities.

{\em The Fr\'echet and Weibull EV limits.} Thus far, we have discussed processes with an asymptotic power-law potential. This means EDs of exponential type, so that their EV limits belong to the Gumbel class. However, our observations hold for the other two EV attractors as well. For the Fr\'echet class we observe a behavior similar to the Gumbel case with $\alpha<1$. Namely, due to the force diminishing at infinity, the DS EV distribution agrees with the CS prediction. For the Weibull class we find that the DS EV distribution converges toward the IID prediction, diverging away from the CS limit. Key equations and supporting figures of these results appear in the Appendix, while derivations and additional extensions can be found in the Supplemental Material \cite{Supp}. We thus conjecture that any process with a potential growing superlinearly, i.e., obeying $\lim_{x\to\infty}x/U(x)=0$, will have its EV statistics controlled by the ED IID behavior in the long-time limit.

{\em Summary and conclusions.} We have demonstrated how the difference between discrete and continuous sampling affects the extreme value (EV) distribution of correlated random variables (RV) generated from Langevin paths. For the Ornstein-Uhlenbeck process, we found that there is a crossover at large measurement times to the statistics of independent and identically distributed RVs drawn from the equilibrium distribution, for any nonzero sampling interval. After providing an intuitive explanation for this phenomenon, we showed it holds for a class of potential fields that are strongly binding. We demonstrated that this is not true for the complementary cases, where the EV distribution diverges from that of independent and identically distributed RVs. Lastly, we showed that our findings apply also to the other two classical limits of EVs, Fr\'echet and Weibull, which were studied via two example cases.

The profound sensitivity of the EV theory of correlated continuous processes to the method of sampling suggests that similar effects will be present also in more general models. Further, any changes encountered in the statistics of EVs may be related to the sampling problem found here, and not to a real change in the physical properties of the system, as we explained. Exploring these issues for models such as fractional Brownian motion, continuous time random walks, processes with demographic or multiplicative noise, and statistics of first-passage times of discretely sampled processes remains an open challenge.

\begin{acknowledgments}

{\em Acknowledgments.} The support of the Israel Science Foundation via Grant No. 1614/21 is acknowledged.

\end{acknowledgments}

\setcounter{equation}{0}
\renewcommand{\theequation}{A\arabic{equation}}

{\em Appendix: The Fr\'echet and Weibull EV limits.} We first consider a potential which grows logarithmically for large displacements \cite{Fogedby,Dechant,Hirschberg}, $U(x)=(\beta/2)\ln(1+x^2)$ with $\beta>1$. Here, the Boltzmann-Gibbs ED decays as a power law, hence the IID limit belongs to the Fr\'echet class. Studying the mode, $z_0$, of the EV distribution obtained from this Langevin process, we find that
\begin{equation}
\label{eqA01}
	z_0^{\rm iid} \sim N^{1/(\beta-1)} , \quad z_0^{\rm cs} \sim T^{1/(\beta+1)} .
\end{equation}
Namely, the IID and CS limits in Eq.~(\ref{eqA01}) display different power-law decays (note that $T=N\Delta$). This is evidenced in Fig.~\ref{fig5}, where for large $T$s the CS limit dominates the EV distribution, whereas for small $T$s the IID picture wins. The potential grows at infinity slower than linearly, hence the CS limit describes the EV distribution correctly at long times. See the Supplemental Material \cite{Supp} for the complete derivation leading to Fig.~\ref{fig5}.
\begin{figure}
	\includegraphics[width=0.8\columnwidth]{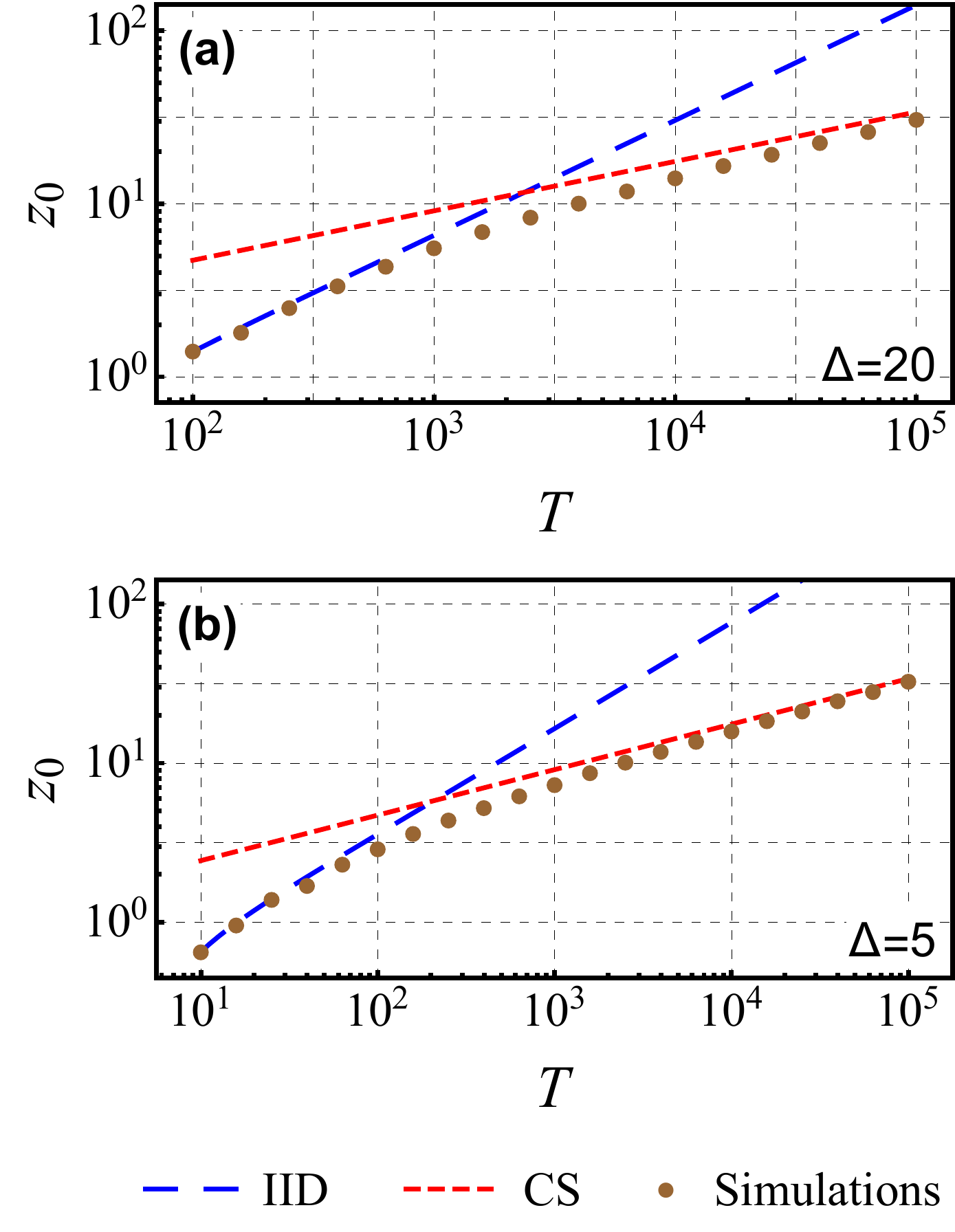}
	\caption{(Color online) {\bf The Fr\'echet class:} The EV mode $z_0$ of a DS Langevin process $x(t)$, which evolves in time according to Eq.~(\ref{eq01}) ($D=1$), but with a deterministic force of $-U'(x)$, where $U(x) = (\beta/2)\ln(1+x^2)$ and $\beta=2.5$, for (a) $\Delta=20$ and (b) $\Delta=5$. Seen are stochastic simulations of the Langevin equation (brown disks), the IID limit (dashed blue line), and the CS limit (short-dashed red line). The CS limit dominates the DS EV distribution for large measurement times due to the force diminishing at $x\to\infty$, while for smaller $T$s the IID limit prevails. Each mode was calculated by maximizing a tenth-order polynomial fitted to a probability density function constructed out of $10^5$ EVs whose initial conditions are $x=0$, obtained using the Euler–Maruyama method with an underlying time increment of $0.01$ and a reflective boundary condition at $x=0$.}
\label{fig5}
\end{figure}

\begin{figure}
	\includegraphics[width=0.8\columnwidth]{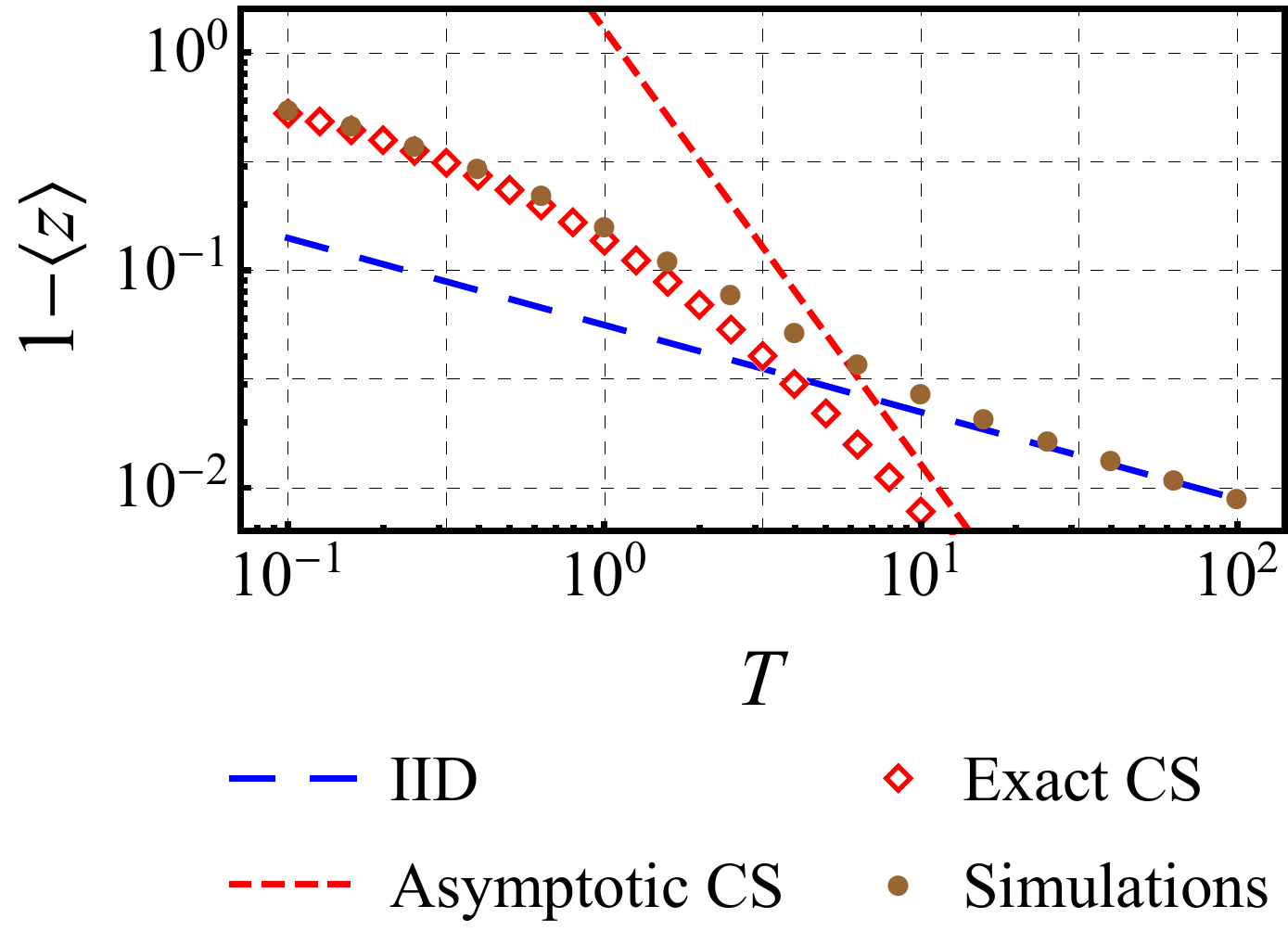}
	\caption{(Color online) {\bf The Weibull class:} The deviation of the mean EV $\langle z \rangle$ from its maximal possible value for a DS Langevin process $x(t)$, which evolves in time according to Eq.~(\ref{eq01}) ($D=1$), but with a deterministic force of $-U'(x)$, where $U(x)$ is given by Eq.~(\ref{eqA02}), $\gamma=2.5$, and $\Delta = 10^{-3}$. Seen are stochastic simulations of the Langevin equation (brown disks), the IID limit (dashed blue line), the long-time asymptotics of the CS limit (short-dashed red curve), and the exact CS limit obtained numerically (hollow red squares). A clear transition from the CS limit to the IID prediction can be observed when the overall measurement time $T$ is increased. Each mean is made of $10^4$ maxima, whose initial conditions are $x=0$, obtained using the Euler–Maruyama method with a varying underlying time increment with a maximal magnitude of $10^{-5}$, and a reflective boundary condition at $x=0$.}
\label{fig6}
\end{figure}
Secondly, we consider a potential corresponding to a particle confined to a finite interval, $x(t)\in[0,1]$,
\begin{equation}
\label{eqA02}
	U(x) = (\gamma-1)\ln\left(\frac{1}{1-x}\right) .
\end{equation}
Note that $\gamma=1$, assuming reflective boundary conditions at $x=0$ and $x=1$, corresponds to a particle freely diffusing in a box. Since here the Boltzmann-Gibbs ED has a finite upper support point, the IID limit belongs to the Weibull class. As the particle's movement is bounded, its maximum value cannot exceed $1$; hence it proves useful to study the quantity $1 - \langle z \rangle$, i.e., the deviation of the mean EV from its maximal possible value. Similarly to the Gumbel case with $\alpha>1$, the CS prediction is entirely off for large measurement times. In general, for the IID limit and any $\gamma>0$, we find the following power-law decay rate:
\begin{equation}
\label{eqA03}
	1-\left<z\right>_{\rm iid} \sim N^{-1/\gamma} .
\end{equation}
However, for the CS limit and $\gamma>2$, we obtain a different power law,
\begin{equation}
\label{eqA04}
	1-\left<z\right>_{\rm cs} \sim T^{-1/(\gamma-2)} ,
\end{equation}
while for $0<\gamma<2$, the decay rate becomes exponential-like. Specifically for a particle freely diffusing in a box, where $\gamma=1$, we obtain
\begin{equation}
\label{eqA05}
	\quad 1-\left<z\right>_{\rm cs} \sim \frac{8}{\pi^3T} \exp\left(-\frac{\pi^2}{4}T\right) .
\end{equation}
To illustrate these results, we first set $\gamma=2.5$ in Fig.~\ref{fig6}. Plotting the deviation of the mean EV from its maximal possible value, $1-\langle z \rangle$, versus the overall measurement time, $T$, we see that even if one takes a small sampling time of $\Delta = 10^{-3}$, the CS limit fails for large $T$, and the IID limit takes control of the EVs, with the ED as an underlying measure. For the other regime, we set $\gamma=1$, giving the example of a particle freely diffusing in a box, as mentioned; see Fig.~\ref{fig7}. It is clear that here too the CS limit fails for large $T$, while the IID limit works excellently. This again marks a qualitative difference between DS with any finite $\Delta$ to the CS limit of $\Delta=0$, here for this example of particles freely diffusing in a box. The complete derivation leading to Figs.~\ref{fig6} and \ref{fig7} can be found in the Supplemental Material \cite{Supp}.
\begin{figure}
	\includegraphics[width=0.8\columnwidth]{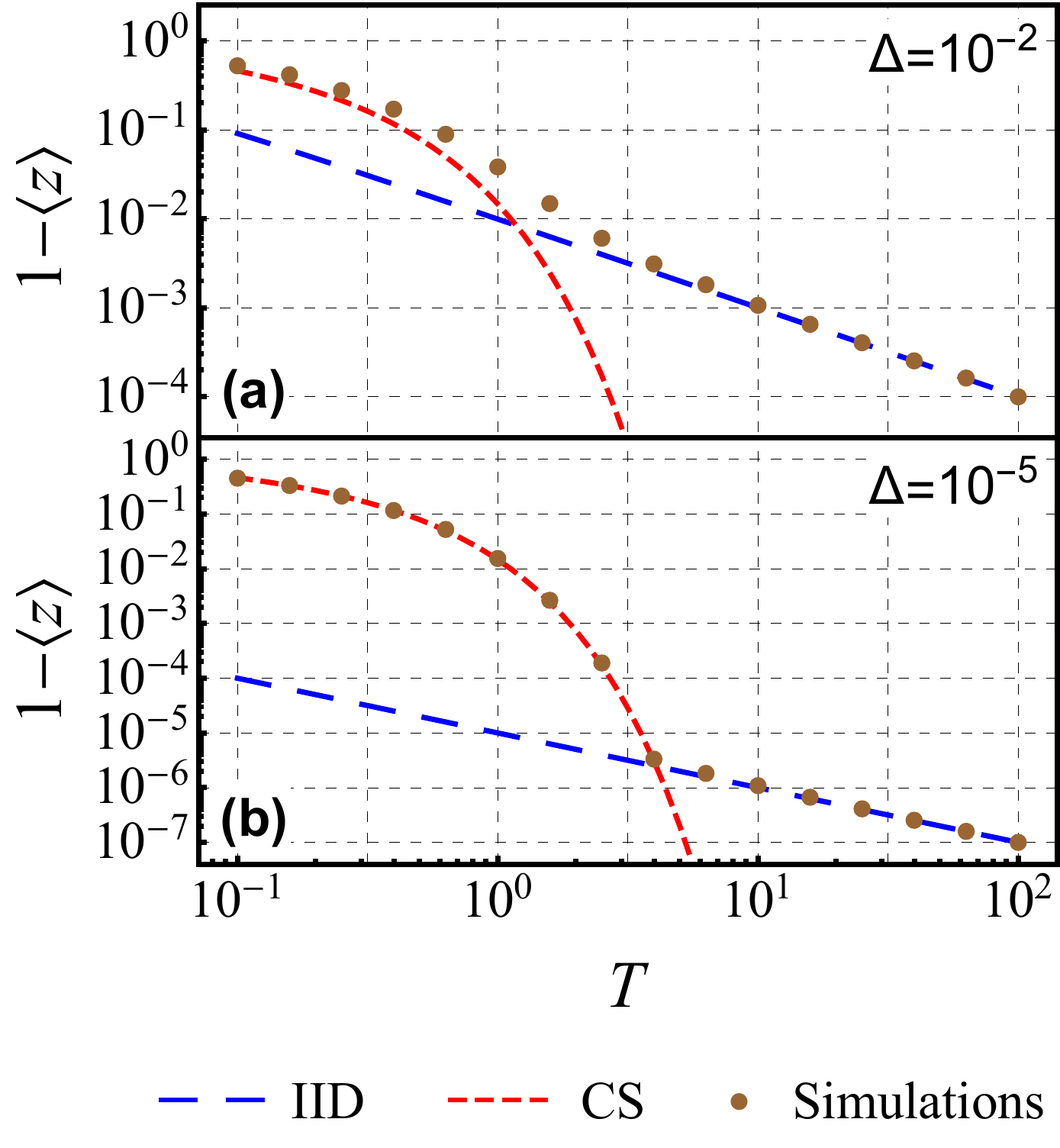}
	\caption{(Color online) {\bf A particle freely diffusing in a box:} The deviation of the mean EV $\langle z \rangle$ from its maximal possible value for a free particle, namely $\gamma=1$ in Eq.~(\ref{eqA02}), confined to $[0,1]$ and controlled by Eq.~(\ref{eq01}) (with $D=1$ and a vanishing deterministic force). Seen are stochastic simulations of the Langevin equation (brown disks), the IID limit (dashed blue line), and the CS limit (short-dashed red curve). (a) For $\Delta = 10^{-2}$, a clear transition from the CS limit to the IID limit can be observed when the overall measurement time $T$ is increased. (b) When decreasing $\Delta$ to $10^{-5}$, the transition occurs at a larger $T$. Each mean is made of $10^4$ maxima whose initial conditions are $x=0$, obtained using the Euler–Maruyama method with an underlying time increment of $10^{-6}$ and reflective boundary conditions at $x=0$ and $x=1$.}
\label{fig7}
\end{figure}

\begin{titlepage}
\title{Supplemental Material for:\\Discrete Sampling of Extreme Events Modifies Their Statistics}
\maketitle
\end{titlepage}

\onecolumngrid
\setcounter{page}{1}
\setcounter{equation}{0}
\renewcommand{\theequation}{SM\arabic{equation}}

\noindent
In what follows, equations and figures that are numbered without the prefix ``SM'' refer to their main text counterparts.

\section{Further details regarding Eq.~(4)}

\subsection{Obtaining the stochastic map}

We start from the rescaled Langevin equation
\begin{equation}
	\frac{\dd}{\dd t}x(t) = -x(t)+\sqrt{2}\eta(t) ,
\end{equation}
where $\eta(\cdot)$ corresponds to standard Gaussian white noise, obeying $\langle\eta(t)\eta(t')\rangle = \delta(t-t')$ and having a zero mean, with $\delta(\cdot)$ denoting the delta function of Dirac. The time $t$ lies in the interval $[0,T]$, where $T$ is the overall measurement duration. With $x(0)$ denoting the initial condition, this equation has the following solution,
\begin{equation}
\label{equation: eigen derivation 1}
	x(t) = e^{-t}x(0) + \sqrt{2}\int_0^{t} \dd t' \, e^{t'-t} \eta(t') ,
\end{equation}
which can be easily verified by differentiation. For some $\Delta>0$, Eq.~(\ref{equation: eigen derivation 1}) can be brought to the following form,
\begin{equation}
\label{equation: eigen derivation 2}
	x(t+\Delta) = e^{-\Delta}x(t) + \sqrt{2}\int_0^{\Delta}\dd t'\, e^{t'-\Delta}\eta(t+t') .
\end{equation}
A discrete sampling (DS) of step $\Delta$ means that one deals with the sequence $\{x_0,x_1,\dots,x_N\}$, where $n \equiv t/\Delta$, $x_n \equiv x(\Delta n)$, and $N \equiv T/\Delta$. Then, we can rewrite Eq.~(\ref{equation: eigen derivation 2}) as
\begin{equation}
\label{equation: eigen derivation 3}
	x_{n+1} = e^{-\Delta}x_n + \eta_n , \quad \eta_n \equiv \sqrt{2}\int_0^{\Delta}\dd t'\, e^{t'-\Delta}\eta(t+t') ,
\end{equation}
where $\eta_n$ is an independent and identically distributed (IID) Gaussian random variable (RV) with zero mean and a variance of $1-\exp(-2\Delta)$.

\subsection{Derivation of Eq.~(4)}

Starting with some distribution for the initial condition $x_0=x(0)$, denoted by $P_0(x)$, the distribution of an $x_n$ obtained after $n$ iterations of Eq.~(\ref{equation: eigen derivation 3}), denoted by $P_n(x)$, satisfies
\begin{equation}
\label{equation: eigen derivation 4}
	P_n(x) \equiv \text{Prob}(x_n=x) = \int_{-\infty}^{\infty}\dd x'\, \text{Prob}\left(x_{n-1}=x' \wedge \eta_{n-1}=x-e^{-\Delta}x'\right) = \int_{-\infty}^{\infty}\dd x'\, P_{n-1}(x') K\left(x-e^{-\Delta}x'\right) ,
\end{equation}
where $K(\cdot)$ is the probability density function (PDF) of the IID RVs $\{\eta_n\}$. Clearly, the DS extreme value (EV) $z_{\rm ds} \equiv \max_{0\le n\le N}(x_n)$ is smaller than $z$ if and only if all the $x_n$s, $0 \le n \le N$, are less than $z$. The recurrence equation governing this event is obtained by replacing $P_n(x) \to \theta(z-x)P_n(x;z)$ in Eq.~(\ref{equation: eigen derivation 4}), where $\theta(\cdot)$ is the Heaviside step function, yielding
\begin{equation}
\label{equation: eigen derivation 5}
	P_n(x;z) = \theta(z-x) \int_{-\infty}^z\dd x'\, P_{n-1}(x';z) K\left(x-e^{-\Delta}x'\right) .
\end{equation}
Equation~(\ref{equation: eigen derivation 5}) is a linear map from $P_{n-1}(x;z)$ to $P_n(x;z)$, and so it is solvable by an eigenvalue expansion, $P_n(x;z) = \sum_{\Lambda} A_{\Lambda}(z) \Lambda^n(z) P_{\Lambda}(x;z)$. Due to the cutoff at $z$, probability is lost in each iteration, and the eigenvalues are all smaller than unity. Thus, for large $n$, the expansion is dominated by the ground state eigenvalue $\Lambda_*(z)$ and eigenfunction $P_*(x;z)$, which obey,
\begin{equation}
\label{equation: eigen derivation 6}
	\Lambda_*(z)P_*(x;z) = \int_{-\infty}^z\dd x'\, P_*(x';z) K\left(x-e^{-\Delta}x'\right) .
\end{equation}
Note that redefining the domain of $x$ to be $(-\infty,z]$ allows us to discard the Heaviside step function. The kernel of Eq.~(\ref{equation: eigen derivation 6}) is simply the PDF of the RV $\eta_n$, namely a Gaussian distribution with zero mean and a variance of $1-\mu^2$,
\begin{equation}
	K(\xi) = \frac{1}{\sqrt{1-\mu^2}}\phi\left(\frac{\xi}{\sqrt{1-\mu^2}}\right) ,
\end{equation}
where $\mu \equiv \exp(-\Delta)$. Thus, Eq.~(4) is obtained.

\subsection{The $\Delta\to0$ limit}

The long-time behavior of the continuous sampling (CS) limit can be retrieved from Eq.~(4) by taking $\Delta\to 0$, leading to $\mu \simeq 1-\Delta$ and $\Lambda_*(z) \simeq 1-\Delta\lambda_*(z)$. Changing variables to $\chi=(x-\mu x')/\sqrt{1-\mu^2}$ in the integration and expanding for $\Delta\to 0$ gives
\begin{equation}
	P_*(x;z) \simeq P_*(x;z) + \Delta \left\{
	\begin{aligned}
		&\left[1+\lambda_*(z)\right]P_*(x;z)+x\frac{\dd}{\dd x}P_*(x;z)+\frac{\dd^2}{\dd x^2} P_*(x;z) & x<z \\
		&-\frac{1}{2\Delta} P_*(z;z) & x=z
	\end{aligned}
	\right. .
\end{equation}
Hence, $p_*(x;z) \equiv \lim_{\Delta\to0}P_*(x;z)$ satisfies the differential equation
\begin{equation}
	\frac{\dd^2}{\dd x^2} p_*(x;z) + \frac{\dd}{\dd x} \left[p_*(x;z)x\vphantom{\frac{1}{1}}\right] + \lambda_*(z) p_*(x;z) = 0 ,
\end{equation}
with a boundary condition at $x=z$ of $p_*(z;z)=0$, yielding the solution
\begin{equation}
	p_*(x;z) \propto \exp\left(-\frac{x^2}{4}\right)\text{D}_{\lambda_*(z)}(-x) , \quad \text{D}_{\lambda_*(z)}(-z) = 0 ,
\end{equation}
exactly as in Ref.~\cite{MajumdarSM}. A derivation of the long-time asymptotics of the CS limit for a general Langevin potential field appears below (third section).

\subsection{Numerical solutions}

Numerically, it proves useful to work with representations of the eigenfunction and eigenvalue that are based on their large-$z$ asymptotics. Therefore, we start by making the following substitution,
\begin{equation}
	\tilde{P}_*(x;z) \equiv 1-\frac{P_*(x;z)}{P_*(x;\infty)} , \quad \tilde{\Lambda}_*(z) \equiv 1-\Lambda_*(z) ,
\end{equation}
where $P_*(x;\infty)\equiv\lim_{z\to\infty}P_*(x;z)$, and of course $\lim_{z\to\infty}\Lambda(z)=1$. The solution at $z\to\infty$ can be found by a Fourier transform of the eigenvalue equation, giving $P_*(x;\infty) = \phi(x)$. The integral eigenvalue equation then becomes
\begin{equation}
\label{equation: numerics 1}
	\left[1-\tilde{\Lambda}_*(z)\right]\phi(x)\left[1-\tilde{P}_*(x;z)\right] = \int_{-\infty}^z\dd x'\, \phi(x') K\left(x-\mu x'\right) - \int_{-\infty}^z\dd x'\, \phi(x') \tilde{P}_*(x';z) K\left(x-\mu x'\right) ,
\end{equation}
where the left integral can be performed analytically. Next, we note that taking $z\to\infty$ has a similar mathematical consequence as having $x\to-\infty$. Hence, let us assume that for some negative $x_{\rm m}$ with $|x_{\rm m}| \gg 1$, we can write an iterative approximation for the solution of Eq.~(\ref{equation: numerics 1}),
\begin{equation}
	\tilde{P}_*(x;z) \approx \left\{
	\begin{aligned}
		& \tilde{P}^*_n(x;z) & x_{\rm m}\le x\le z \\
		& 0 & -\infty<x<x_{\rm m}
	\end{aligned}
	\right. ,
\end{equation}
where $\tilde{P}^*_n(x;z)$ is the $[x_{\rm m},z]$-part of the eigenfunction corresponding to the $n$th iteration. Similarly, we denote $\tilde{\Lambda}^*_n(z)$ as the $n$th iteration's eigenvalue. Thus, for $x\in[x_{\rm m},z]$, Eq.~(\ref{equation: numerics 1}) changes to
\begin{equation}
\label{equation: numerics 2}
	\left[1-\tilde{\Lambda}^*_n(z)\right]\phi(x)\left[1-\tilde{P}^*_n(x;z)\right] = \int_{-\infty}^z\dd x'\, \phi(x') K\left(x-\mu x'\right) - \int_{x_{\rm m}}^z\dd x'\, \phi(x') \tilde{P}^*_{n-1}(x';z) K\left(x-\mu x'\right) .
\end{equation}
Assuming $\tilde{P}^*_{n-1}(x;z)$ is known, we discretize $x'$ on the interval $[x_{\rm m},z]$ and calculate the right integral of Eq.~(\ref{equation: numerics 2}). We find $\tilde{\Lambda}^*_n(z)$ by evaluating Eq.~(\ref{equation: numerics 2}) at $x=x_{\rm m}$, where due to continuity $\tilde{P}^*_n(x_{\rm m};z)=0$, yielding
\begin{equation}
\label{equation: numerics 3}
	\left[1-\tilde{\Lambda}^*_n(z)\right]\phi(x) = \int_{-\infty}^z\dd x'\, \phi(x') K\left(x_{\rm m}-\mu x'\right) - \int_{x_{\rm m}}^z\dd x'\, \phi(x')	\tilde{P}^*_{n-1}(x';z) K\left(x_{\rm m}-\mu x'\right) .
\end{equation}
Using this value, we obtain $\tilde{P}^*_n(x;z)$ for $x\in[x_{\rm m},z]$. Starting with $\tilde{P}^*_0(x;z)=0$ and continuing to iterate gives a series of approximations to $\tilde{P}_*(x;z)$ which converges efficiently. The left integral of Eqs.~(\ref{equation: numerics 2}) and (\ref{equation: numerics 3}) can be expressed in a simple closed form, and we get
\begin{align}
	\tilde{\Lambda}^*_n(z) &= \frac{1}{2} \erfc\left[ \frac{z - \mu x_{\rm m}}{\sqrt{2(1-\mu^2)}} \right] + \int_{x_{\rm m}}^z \dd x' \, \frac{\tilde{P}^*_{n-1}(x';z)}{\sqrt{2\pi(1-\mu^2)}} \exp\left[-\frac{(x'-\mu x_{\rm m})^2}{2(1-\mu^2)} \right] , \\
	\tilde{P}^*_n(x;z) &= \left[1-\tilde{\Lambda}^*_n(z)\right]^{-1} \left\{ \frac{1}{2} \erfc\left[ \frac{z - \mu x}{\sqrt{2(1-\mu^2)}} \right] + \int_{x_{\rm m}}^z \dd x' \, \frac{\tilde{P}^*_{n-1}(x';z)}{\sqrt{2\pi(1-\mu^2)}} \exp\left[-\frac{(x'-\mu x)^2}{2(1-\mu^2)} \right] - \tilde{\Lambda}^*_n(z) \right\} . \nonumber
\end{align}
Lastly, we define a measure of convergence to determine the stopping point of this iterative process,
\begin{equation}
	\mathcal{E} \equiv \left|\frac{\tilde{\Lambda}^*_{100m}(z)}{\tilde{\Lambda}^*_{100(m-1)}(z)}-1\right| , \quad 1<m\in\mathbb{N} .
\end{equation}
This prescription was used to obtain the numerical data for $\Lambda_*(z)$ presented in Figs.~2 and 3. The discretization step in $x$ was $0.01$, and we used $x_{\rm m}=-5$ and $\mathcal{E}=10^{-7}$.

\section{Derivation of Eqs.~(7) and (8)}

We start by writing that for large-$z$
\begin{equation}
\label{equation: lambda derivation 1}
	P_*(x;z) \simeq \phi(x)\left\{1 + \phi(z){\cal P}_1(x;z) + [\phi(z)]^2{\cal P}_2(x;z)\right\} , \quad \Lambda_*(z) \simeq 1 - \phi(z)\Lambda_1(z) + [\phi(z)]^2\Lambda_2(z) .
\end{equation}
These expansions are to be understood in the context of a fixed $0\le\mu<1$.

\subsection{Finding the first-order correction}

Plugging the above expansion into Eq.~(4), we get to first order
\begin{equation}
	\phi(x)\left[1 + \phi(z){\cal P}_1(x;z) - \phi(z)\Lambda_1(z)\right] = \int_{-\infty}^z \dd x' \, \phi(x') \frac{1+\phi(z){\cal P}_1(x';z)}{\sqrt{2\pi(1-\mu^2)}}\exp\left[-\frac{(x-\mu x')^2}{2(1-\mu^2)}\right] .
\end{equation}
The zeroth-order equation is satisfied since
\begin{equation}
	\int_{-\infty}^z \dd x' \, \phi(x')\frac{1}{\sqrt{2\pi(1-\mu^2)}}\exp\left[-\frac{(x-\mu x')^2}{2(1-\mu^2)}\right]=\phi(x)\left\{1-\frac{1}{2}\erfc\left[\frac{z-x\mu}{\sqrt{2(1-\mu^2)}}\right]\right\} ,
\end{equation}
where $\erfc(\cdot)$ is the complementary error function. Using the following expansion \cite{Hermite1} of the Gaussian kernel function of Eq.~(4), which holds for $0\le\mu<1$,
\begin{equation}
\label{equation: hermite expansion 1}
	\frac{1}{\sqrt{2\pi(1-\mu^2)}}\exp\left[-\frac{(x-x'\mu)^2}{2(1-\mu^2)}\right] = \phi(x) \sum_{n=0}^{\infty} \frac{\mu^n}{n!}\He_n(x)\He_n(x') ,
\end{equation}
where $\He_n(\cdot)$ is the $n$th probabilists' Hermite polynomial, we obtain to first order
\begin{equation}
\label{equation: lambda derivation 2}
	{\cal P}_1(x;z) - \Lambda_1(z) + \frac{1}{2\phi(z)}\erfc\left[\frac{z-x\mu}{\sqrt{2(1-\mu^2)}} \right] - \int_{-\infty}^{\infty}\dd x'\,\phi(x'){\cal P}_1(x';z)\sum_{n=0}^{\infty} \frac{\mu^n}{n!}\He_n(x)\He_n(x') = 0 .
\end{equation}
Note that we have extended the integral's boundary to infinity, dropping a higher-order correction to be accounted for during the second-order calculation. Exploiting another expansion \cite{Hermite2} similar to the one above,
\begin{equation}
\label{equation: hermite expansion 2}
	\erfc\left[\frac{z-x\mu}{\sqrt{2(1-\mu^2)}}\right] = \erfc\left(\frac{z}{\sqrt{2}}\right) + 2\phi(z) \sum_{n=1}^{\infty} \frac{\mu^n}{n!}\He_n(x)\He_{n-1}(z) ,
\end{equation}
together with expressing the first functional correction as a sum over Hermite polynomials in $x$,
\begin{equation}
	{\cal P}_1(x;z) = \sum_{n=0}^{\infty} c_n(z) \He_n(x) ,
\end{equation}
and using their orthogonality condition (where $\delta_{n,m}$ is the Kronecker delta),
\begin{equation}
	\int_{-\infty}^{\infty} \dd x' \phi(x')\He_{n}(x')\He_m(x') = \delta_{n,m} n! ,
\end{equation}
we get for the first-order expansion
\begin{equation}
	\left[c_0(z)\left(1-\mu^0\right)-\Lambda_1(z)+\frac{1}{2\phi(z)}\erfc\left(\frac{z}{\sqrt{2}}\right)\right]\He_0(x) + \sum_{n=1}^{\infty}\left[c_n(z)\left(1-\mu^n\right)+ \frac{\mu^n}{n!}\He_{n-1}(z)\right]\He_n(x) = 0 .
\end{equation}
Thus, we obtain
\begin{equation}
\label{equation: lambda derivation 3}
	\Lambda_1(z) = \frac{\erfc(z/\sqrt{2})}{2\phi(z)} , \quad c_n(z) = -\frac{\mu^n}{n!}\frac{\He_{n-1}(z)}{1-\mu^n} , \quad n>0 ,
\end{equation}
where $\Lambda_1(z) \sim z^{-1}$ for $z\to\infty$. The value of $c_0(z)$ can be found from the condition ${\cal P}_1(0;z)=0$, since an $x$-independent addition to ${\cal P}_1(x;z)$ is just a change of normalization. This yields
\begin{equation}
	c_0(z) = -\sum_{n=1}^{\infty} c_n(z) \He_n(0) = -\sum_{n=1}^{\infty} \frac{\sqrt{\pi}2^{n/2}c_n(z)}{\Gamma[(1-n)/2]} ,
\end{equation}
where $\Gamma(\cdot)$ is the gamma function.

\subsection{Obtaining the second-order correction}

Since we have an exact solution of the first-order equation, we can move on to the second order. We further expand Eq.~(4) to second-order, obtaining
\begin{align}
\label{equation: lambda derivation 4}
	{\cal P}_2(x;z)-{\cal P}_1(x;z)\Lambda_1(z)+\Lambda_2(z) + \frac{1}{\phi(z)} &\int_z^{\infty} \dd x'\,\phi(x'){\cal P}_1(x';z) \sum_{n=0}^{\infty} \frac{\mu^n}{n!}\He_n(x)\He_n(x') \nonumber \\
	- &\int_{-\infty}^{\infty} \dd x'\,\phi(x'){\cal P}_2(x';z) \sum_{n=0}^{\infty} \frac{\mu^n}{n!}\He_n(x)\He_n(x') = 0 .
\end{align}
The first integral term is the higher-order correction that was dropped in Eq.~(\ref{equation: lambda derivation 2}). As done above, the boundary of the second integral term was extended to infinity (since the contribution from $x'>z$ only enters the calculation of the third-order correction). Let us express ${\cal P}_2(x;z)$ similarly to its first-order counterpart,
\begin{equation}
	{\cal P}_2(x;z) = \sum_{n=0}^{\infty} d_n(z) \He_n(x) .
\end{equation}
Plugging this and Eq.~(\ref{equation: lambda derivation 3}) into Eq.~(\ref{equation: lambda derivation 4}) and rearranging, we get from the terms which multiply $\He_0(x)$ that
\begin{equation}
	\Lambda_2(z) = \sum_{n=1}^{\infty} \frac{\mu^n/n!}{1-\mu^n} \phi_{n,0}(z)\He_{n-1}(z) ,
\end{equation}
with \cite{Hermite3}
\begin{equation}
\label{equation: hermite expansion 3}
	\phi_{n,m}(z) \equiv \frac{1}{\phi(z)} \int_z^{\infty}\dd x\,\phi(x) \He_n(x)\He_m(x) = \sum_{l=0}^L l!\binom{n}{l}\binom{m}{l} \He_{n+m-2l-1}(z) + \delta_{n,m}n! \frac{\erfc(z/\sqrt{2})}{2\phi(z)} ,
\end{equation}
where $L \equiv \min(n,m)-\delta_{n,m}$ and we used the standard convention that a summation from $0$ to $-1$ vanishes. Taking $m=0$ and $n\ge1$, we get $\phi_{n,0}(z)=\He_{n-1}(z)$, hence
\begin{equation}
	\Lambda_2(z) = \sum_{n=1}^{\infty} \frac{\mu^n/n!}{1-\mu^n} \He_{n-1}^2(z) ,
\end{equation}
which, together with $\Lambda_1(z)$ from Eq.~(\ref{equation: lambda derivation 3}), yields Eq.~(7) when plugged into Eq.~(\ref{equation: lambda derivation 1}). Finally, using the identity
\begin{equation}
	\sum_{n=0}^{\infty} \frac{\nu^n}{n!} \He_n^2(z) = \frac{1}{\sqrt{1-\nu^2}} \exp\left(\frac{z^2\nu}{1+\nu}\right) ,
\end{equation}
which arises in the calculation of the density of states of the finite temperature quantum harmonic oscillator \cite{BondarevSM}, we find
\begin{equation}
	\Lambda_2(z) = \sum_{n=1}^{\infty} \int_0^{\mu^n} \frac{\dd\nu}{\sqrt{1-\nu^2}} \exp\left(\frac{z^2\nu}{1+\nu}\right) \mathop{\sim}_{z\to\infty} \frac{(1+\mu)^2}{\sqrt{1-\mu^2}}z^{-2} \exp\left(\frac{z^2\mu}{1+\mu}\right) .
\end{equation}
Together with the asymptotic behavior of $\Lambda_1(z)$ at infinity, the top row of Eq.~(8) is obtained.

\section{Langevin processes with non-linear forces}

Here we provide further details regarding the generalization of our findings to different Langevin equations. We start with generalizing Eq.~(1),
\begin{equation}
\label{equation: additional 1}
	\frac{\dd}{\dd t} x(t) = - \frac{D}{\kb\TT}U'[x(t)] + \sqrt{2D}\eta(t) ,
\end{equation}
where $U(x)$ is the potential, $D$ is the diffusion coefficient, $\kb$ is Boltzmann's constant, $\TT$ is the temperature, $\eta(t)$ is the same standard Gaussian white noise as defined above, and primes denote derivatives. The equilibrium distribution of this process is given by the Boltzmann-Gibbs measure,
\begin{equation}
\label{equation: additional 2}
	P_{\rm eq}(x) = \frac{1}{Z}\exp\left[-\frac{U(x)}{\kb\TT}\right] , \quad Z \equiv \int_{-\infty}^{\infty}\dd x \exp\left[-\frac{U(x)}{\kb\TT}\right] ,
\end{equation}
where $Z$ is the partition function. Therefore, the IID limit for the EV cumulative distribution function (CDF) is
\begin{equation}
\label{equation: additional 3}
	F^{\rm iid}_N(z) \equiv \lim_{\Delta\to\infty}\text{Prob}\left(z_{\rm ds}<z\right) = \left[ \int_{-\infty}^z \dd x\, P_{\rm eq}(x) \right]^N .
\end{equation}

On the other hand, the CS limit is obtainable by generalizing the method described in Ref.~\cite{MajumdarSM} for the Ornstein-Uhlenbeck (OU) process to go beyond this specific model. The main idea is to solve the Fokker-Planck representation of Eq.~(\ref{equation: additional 1}),
\begin{equation}
\label{equation: additional 4}
	\frac{\partial}{\partial t}p(x,t|z) = D\frac{\partial^2}{\partial x^2}p(x,t|z) + \frac{D}{\kb\TT}\frac{\partial}{\partial x} \left[ p(x,t|z) U'(x) \vphantom{\frac{1}{1}} \right] ,
\end{equation}
with an initial condition of $p(x,0|z)=\delta(x)$ and boundary conditions of $p(-\infty,t|z)=p(z,t|z)=0$. Here, the notation $p(x,t|z)$ denotes the probability for a particle described by $x(t)$ to arrive at $x$ on time $t$, while always staying below the value $z$. The EV CDF at the total measurement time $T$ is then
\begin{equation}
	F^{\rm cs}_T(z) \equiv \lim_{\Delta\to0}\text{Prob}\left(z_{\rm ds}<z\right) = \text{Prob}\left(z_{\rm cs}<z\right) = \int_{-\infty}^z \dd x\, p(x,T|z) .
\end{equation}
Formally, one can write a solution via an eigenvalue expansion,
\begin{equation}
	p(x,t|z) = \sum_{\lambda} A_{\lambda}(z)e^{-D\lambda(z)t}p_{\lambda}(x;z) ,
\end{equation}
where $\{\lambda(z)\}$ are the eigenvalues and $\{p_{\lambda}(x;z)\}$ are the eigenfunctions, which obey
\begin{equation}
\label{equation: additional 5}
	\frac{\dd^2}{\dd x^2} p_{\lambda}(x;z) + \frac{\dd}{\dd x}\left[p_{\lambda}(x;z) \frac{U'(x)}{\kb\TT} \right] + \lambda(z)p_{\lambda}(x;z) = 0 ,
\end{equation}
with boundary conditions of $p_{\lambda}(-\infty;z)=p_{\lambda}(z;z)=0$. Let us denote as $\lambda_*(z)$ and $p_*(x;z)$ the smallest eigenvalue and its associated eigenfunction. Note that for $z\to\infty$ we have $\lambda_*(z)\to0$, which is to be expected since in this limit the boundary conditions of $p_*(x;z)$ yield the equilibrium density $P_{\rm eq}(x)$, associated with a zero eigenvalue. Thus, in the long measurement time limit $t=T\to\infty$, we can replace $p_*(x;z)$ with $P_{\rm eq}(x)$ and set $A_*=1$ (due to the initial condition being normalized), which yields
\begin{equation}
\label{equation: additional 6}
	F^{\rm cs}_T(z) \sim \exp\left[-D\lambda_*(z)T\right] ,
\end{equation}
up to exponentially small corrections.

Below we consider three example cases, classified according to the large $N$ limit of Eq.~(\ref{equation: additional 3}). Our aim is to find the IID and CS representations of the EV distribution of the DS process for the three EV basins of attraction. Using these limiting functions, we verify the general nature of our study.

\subsection{The Gumbel class}

The Gumbel class occurs when $-\ln[P_{\rm eq}(x)] \propto x^{\alpha}$ for $x\to\infty$ and $\alpha>0$. Accordingly, here we assume a potential with an asymptotic behavior of
\begin{equation}
\label{equation: gumbel 1}
	U(x) \propto |x|^{\alpha} , \quad \alpha>0 ,
\end{equation}
for $x\to\pm\infty$. As mentioned in the main text, the DS EV distribution displays two different behaviors depending on the value of $\alpha$, and therefore, we split our derivation of the CS limit into two. Starting with the case $\alpha>1$, we solve Eq.~(\ref{equation: additional 5}) for $x\in(-\infty,z]$ with boundary conditions of $p_*(-\infty;z)=0$ and $p_*(z;z)=0$, from which $\lambda_*(z)$ emerges. We do so by using perturbation theory around $\lambda_*(z)=0$, for which $z\to\infty$ as mentioned above. Writing $p_*(x)=p^*_0(x)+p^*_1(x;z)$, Eq.~(\ref{equation: additional 5}) reads for the zeroth order
\begin{equation}
	\frac{\dd^2}{\dd x^2} p^*_0(x) + \frac{\dd}{\dd x}\left[p^*_0(x) \frac{U'(x)}{\kb\TT} \right] = 0 ,
\end{equation}
whose general solution is
\begin{equation}
	p^*_0(x) = C_1 y_1(x) + C_2 y_2(x)
\end{equation}
where
\begin{equation}
	y_1(x) \equiv \exp\left[-\frac{U(x)}{\kb\TT}\right] , \quad y_2(x) \equiv y_1(x) \int_0^x \dd\xi \exp\left[\frac{U(\xi)}{\kb\TT}\right] ,
\end{equation}
with boundary conditions of $p^*_0(-\infty)=p^*_0(\infty)=0$. Since $y_2(x)$ decays algebraically when $x\to\pm\infty$, more precisely $y_2(x) \propto |x|^{1-\alpha}$, it needs to be discarded, as the solution should approach zero for $x\to-\infty$ in an exponential manner. Therefore, we have
\begin{equation}
	p^*_0(x) = C_1 y_1(x)
\end{equation}
as the zero-order solution. For the first order, we obtain the inhomogeneous equation
\begin{equation}
\label{equation: gumbel 2}
	\frac{\dd^2}{\dd x^2}p^*_1(x;z) + \frac{\dd}{\dd x} \left[ p^*_1(x;z) \frac{U'(x)}{\kb\TT} \right] = -\lambda_*(z) p^*_0(x) .
\end{equation}
By the method of variation of parameters, the general solution of Eq.~(\ref{equation: gumbel 2}) is given by
\begin{equation}
\label{equation: gumbel 3}
	p^*_1(x;z) = C_1 \lambda_*(z) \left[ y_1(x)\int_{\xi_1}^x\dd\xi\,y_2(\xi) - y_2(x)\int_{\xi_2}^x\dd\xi\,y_1(\xi) \right] ,
\end{equation}
where $\xi_1$ and $\xi_2$ are arbitrary constants. As mentioned, the decay at $x\to-\infty$ should be exponential, hence the coefficient of $y_2(x)$ must vanish in this limit. Thus, we must choose $\xi_2=-\infty$, which gives
\begin{equation}
	p_*(x;z) \simeq C_1 y_1(x) + C_1 \lambda_*(z) \left[ y_1(x) \int_{\xi_1}^x \dd\xi\, y_2(\xi) - y_2(x) \int_{-\infty}^x \dd\xi\, y_1(\xi) \right] .
\end{equation}
Setting this to zero at $x=z$ yields
\begin{equation}
\label{equation: gumbel 4}
	\lambda_*(z) \simeq \frac{1}{Z} \left\{ \int_0^z \dd x\, \exp\left[\frac{U(x)}{\kb\TT}\right] \right\}^{-1} ,
\end{equation}
where $Z$ is the partition function defined in Eq.~(\ref{equation: additional 2}). Further approximating this for $z\to\infty$, we find
\begin{equation}
	\lambda_*(z) \sim \frac{1}{Z} \frac{U'(z)}{\kb\TT} \exp\left[-\frac{U(z)}{\kb\TT}\right] .
\end{equation}

Things are more complicated when the potential grows slower than linearly, i.e. $0<\alpha<1$ in Eq.~(\ref{equation: gumbel 1}), so that the force decays to zero for large $x$. Here, the spectrum of the Fokker-Planck equation on the semi-infinite domain $-\infty<x\le z$ is not discrete, and the eigenvalues go continuously to $0$. Treating this case requires a very different approach, which is beyond the scope of this paper. However, if we use a reflective boundary condition at $x=0$, solving the problem of $x\in[0,z]$ instead, the spectrum is indeed discrete and we can proceed as before. Therefore, we now solve Eq.~(\ref{equation: additional 5}) over the domain $x\in[0,z]$, where the boundary conditions are $\dd p_*(x;z)/\dd x|_{x=0}=p_*(z;z)=0$. The zero-order has the same general solution, and its boundary conditions read $p^{*\prime}_0(0)=p^*_0(\infty)=0$. Since $y_2'(0)=1$, we have the same solution for the zeroth-order. Hence, we obtain the same inhomogeneous equation to first-order, solved via the method of variation of parameters to yield Eq.~(\ref{equation: gumbel 3}). However, this time due to the boundary condition at $0$ and given that $y_2'(0)=1$, we must choose $\xi_2=0$, and we obtain an eigenvalue which is twice the magnitude of the $\alpha>1$ case, i.e. $2\lambda_*(z)$.

Note that for large $z$, the effective IID underlying CDF,
\begin{equation}
	1-F^{\rm iid}(z) = \int_z^{\infty} \dd x\, P_{\rm eq}(x) \propto \frac{1}{U'(z)}\exp\left[-\frac{U(z)}{\kb\TT}\right] ,
\end{equation}
differs from the CS effective CDF,
\begin{equation}
	1-F^{\rm cs}(z) \propto \lambda_*(z) \propto U'(z) \exp\left[-\frac{U(z)}{\kb\TT}\right] ,
\end{equation}
by a prefactor $\propto[U'(z)]^2$, proportional to $z^{2\alpha-2}$ as $z\to\infty$. Thus, for $0<\alpha<1$, the latter PDF decays faster than the former, which means that the average EV in the IID picture is larger than the CS one for large $T$. As $z_{\rm ds} \le z_{\rm cs}$ dictates that $\langle z_{\rm ds} \rangle \le \langle z_{\rm cs} \rangle$, one must infer that for forces which vanish at large distance, the asymptotic behavior at large-$z$ is bounded from above by the CS limit, and hence it cannot approach the IID limit, in contradistinction to what happens for diverging forces, e.g. the OU model. For the case of $\alpha=1$, namely an asymptotically linear potential, the $z^{2\alpha-2}$ prefactor is absent. In this case, both the effective IID EV distribution and its CS limit counterpart are asymptotically purely exponential. This linear potential case was discovered to be marginal also for other problems which are not related to DS, see for example Ref.~\cite{SanjibSM}, where the authors find a freezing transition in the long-time decay rate of the first-passage distribution of a particle whose trajectory is controlled by Eq.~(\ref{equation: additional 1}). Lastly, we stress that due to the exponential-like decay of the IID underlying CDF and of the CS effective CDF, both limits belong to the Gumbel class for any $\alpha>0$, hence the associated DS processes are probably of an identical nature.

The above results are demonstrated in Fig.~4, where we assumed the following shape for the potential,
\begin{equation}
	U(x) = \frac{1}{\alpha}(1+x^2)^{\alpha/2}
\end{equation}
with $D=\kb\TT=1$, which has the same asymptotics as Eq.~(\ref{equation: gumbel 1}). The IID curves were computed from Eq.~(\ref{equation: additional 3}), while the CS curves from Eq.~(\ref{equation: additional 6}), with $\lambda_*(z)$ given by Eq.~(\ref{equation: gumbel 4}). For the OU model of $\alpha=2$ we used the exact solution for $\lambda_*(z)$, given using the parabolic cylinder function just above Eq.~(5) in the main text. When $\alpha<1$, we accounted for the reflection at $x=0$ by replacing $Z\to Z/2$ in the IID and CS formulas, Eqs.~(\ref{equation: additional 3}) and (\ref{equation: gumbel 4}), respectively.

\subsection{The Fr\'echet class}

The Fr\'echet class occurs when $P_{\rm eq}(x) \propto x^{-\beta}$ for $x\to\infty$ and $\beta>1$. We assume the following large-$|x|$ behavior for the potential,
\begin{equation}
\label{equation: frechet 1}
	U(x) \sim U_{\infty}\ln(|x|/a) , \quad U_{\infty}>0 , \quad a>0 ,
\end{equation}
where $\beta \equiv U_{\infty}/(\kb\TT)>1$, such that for large enough $z$ we have for the EV IID limit,
\begin{equation}
	F^{\rm iid}_N(z) \simeq \left[1 - \frac{1}{Z}\int_z^{\infty} \dd x \left(\frac{a}{x}\right)^{\beta} \right]^N = \left[ 1 - \frac{a^{\beta}z^{1-\beta}}{(\beta-1)Z} \right]^N ,
\end{equation}
where $Z$ is the partition function defined in Eq.~(\ref{equation: additional 2}). Note that depending on $\beta$, the mean of this distribution is not always well-defined. Hence, here observe the mode $z_0$ of the EV PDF. For single maximum PDFs, it is defined as
\begin{equation}
\label{equation: frechet 2}
	\left. \frac{\dd}{\dd z} \text{PDF}(z) \right|_{z=z_0} = 0 .
\end{equation}
Thus, when $N$ is large we obtain the EV mode in the IID limit,
\begin{equation}
	z_0^{\rm iid}(N) \simeq a \left[ \frac{a}{Z} \frac{N(\beta-1)+1}{\beta(\beta-1)} \right]^{1/(\beta-1)} .
\end{equation}

We now move to the CS limit. In Ref.~\cite{DechantSM}, the authors presented an approximation to $p_{\lambda}(x;z)$ with $z\to\infty$ for a potential behaving as Eq.~(\ref{equation: frechet 1}). By moving to the Schr\"odinger representation,
\begin{equation}
	\psi_k(x;z) \equiv \frac{p_{\lambda}(x;z)}{\sqrt{P_{\rm eq}(x)}} , \quad k \equiv \sqrt{\lambda} ,
\end{equation}
they found that in the large-$x$ regime, up to a normalization constant
\begin{equation}
	\psi_k(x;z\to\infty) \simeq \sqrt{\frac{x}{a}} \left[- \Gamma(\nu)\frac{Z}{a} \left(\frac{ka}{2}\right)^{2-\nu} \text{J}_{\nu}(kx) + \Gamma(1-\nu)\left(\frac{ka}{2}\right)^{\nu} \text{J}_{-\nu}(kx) \right] , \quad \nu \equiv \frac{\beta+1}{2} ,
\end{equation}
where $\text{J}_{\cdot}(\cdot)$ is Bessel's function of the first kind, see Eqs.~(38), (40), and (50) of Ref.~\cite{DechantSM}. To obtain the solution for a finite $z$, here we need to change the upper boundary condition from $x=\infty$ to $x=z$. This can be done simply by demanding that $\psi_{k}(z;z\to\infty)=0$ for a certain $k_*(z)$. Using the small argument expansion of the Bessel function, $J_{\nu}(\xi) \sim (\xi/2)^{\nu}/\Gamma(\nu+1)$, we obtain
\begin{equation}
	\lambda_*(z) = k_*^2(z) \simeq \frac{4\nu}{a^2}\frac{a}{Z}\left(\frac{a}{z}\right)^{2\nu} .
\end{equation}
Plugging this into Eq.~(\ref{equation: additional 6}) yields the CS EV CDF, which for large $z$ behaves as
\begin{equation}
	F^{\rm cs}_T(z) \simeq \left[1-2(\beta+1)\frac{a}{Z}\left(\frac{a}{z}\right)^{\beta+1} \right]^{DT/a^2} ,
\end{equation}
from which the mode is found for large $T$,
\begin{equation}
\label{equation: frechet 3}
	z_0^{\rm cs}(T) \simeq a \left[ \frac{(1+\beta)^2}{1+\beta/2}\frac{DT}{aZ} \right]^{1/(\beta+1)} .
\end{equation}

Note that the derivation presented in Ref.~\cite{DechantSM} assumes an even wave function for the perturbative solution of the small-$x$ inner region. Consequently, the above expressions are valid when a reflective boundary condition at the origin is assumed, i.e. $0\le x \le z$, as with the $0<\alpha<1$ case of the Gumbel domain. A semi-infinite case of $x\in(-\infty,z]$ can be treated by generalizing the results obtained in Ref.~\cite{DechantSM} for the solution of $\psi_k(x;z\to\infty)$ in the small-$x$ regime for a general parity wave function, which is again beyond the scope of this paper. Also, we see that the IID underlying CDF, $1-F^{\rm iid}(z) \propto z^{-\beta+1}$, differs from the CS effective CDF, $1-F^{\rm cs}(z) \propto z^{-\beta-1}$, by a prefactor $\propto z^{-2}$. Thus, for any $\beta>1$, the latter PDF decays faster than the former, and by the same argument made for processes that belong to the Gumbel class with $0<\alpha<1$, one infers that DS processes which belong to the Fr\'echet class cannot converge to their respective IID limits when $T$ becomes large, in contradistinction to processes which belong to the Gumbel class with $\alpha>1$, or to processes attracted to the Weibull class (for the latter, see below). One can also see that both the IID and CS limits above lie in the Fr\'echet domain, as the respective CDFs decay as a power-law.

We put these predictions to a test using the potential
\begin{equation}
\label{equation: frechet 4}
	U(x) = \frac{\beta}{2}\ln\left(1+x^2\right) ,
\end{equation}
which for large $x$ behaves as Eq.~(\ref{equation: frechet 1}) with $a=1$. Setting $\beta=2.5$ and $D=1$, we find that for large $T$s the CS limit dominates the EV distribution, whereas for smaller $T$s the IID picture wins, as depicted in Fig.~5. There, the IID limit is given by solving Eq.~(\ref{equation: frechet 2}), and the CS limit is given by Eq.~(\ref{equation: frechet 3}). We accounted for the reflection at $x=0$ by setting $Z \to Z/2$ in the IID expression.

\subsection{The Weibull class}

The Weibull class occurs when there is a finite upper bound $L$ on the interval in which the particle is allowed to travel and $P_{\rm eq}(x)$ decays slowly enough for $x\to L$. As the particle's movement is bounded, its maximum value cannot exceed $L$, hence it proves useful to study the quantity $L - \langle z \rangle$, i.e., the deviation of the mean EV from its maximal possible value. Here, we consider the following example case,
\begin{equation}
\label{equation: weibull 1}
	U(x) = \left\{
	\begin{aligned}
		& U_0\ln\left(\frac{L}{L-x}\right) & 0 \le x \le L \\
		& \infty \vphantom{\frac{1}{1}} & \text{otherwise}
	\end{aligned}
	\right. ,
\end{equation}
such that $L>0$ and $\gamma \equiv 1+U_0/(\kb\TT) > 0$. Note that $\gamma=1$ corresponds to $U_0=0$, i.e. to a particle freely diffusing in a box, which is solved below. For now, let us write down the IID equilibrium measure,
\begin{equation}
	P_{\rm eq}(x) = \frac{\gamma}{L} \left(1-\frac{x}{L}\right)^{\gamma-1} ,
\end{equation}
which can be rescaled as $Q_{\rm eq}(\chi)\dd\chi \equiv P_{\rm eq}(x)\dd x$ with $x=L\chi$, such that $L$ vanishes from the expressions,
\begin{equation}
	Q_{\rm eq}(\chi) = \gamma(1-\chi)^{\gamma-1} , \quad F^{\rm iid}(\chi) = 1-(1-\chi)^{\gamma} .
\end{equation}
The IID limit of the DS EV CDF is thus,
\begin{equation}
	F^{\rm iid}_N(\zeta) = \left[1-(1-\zeta)^{\gamma}\right]^N ,
\end{equation}
where $\zeta\equiv z/L$, which yields
\begin{equation}
\label{equation: weibull 2}
	1-{\langle\zeta\rangle}_{\rm iid} = 1 - \int_0^1 \dd\zeta \left[\frac{\dd}{\dd\zeta} F^{\rm iid}_N(\zeta)\right] \zeta = \frac{\Gamma(1+1/\gamma)\Gamma(1+N)}{\Gamma(1+1/\gamma+N)} \sim \Gamma\left(1+\frac{1}{\gamma}\right) N^{-1/\gamma} = \Gamma\left(1+\frac{1}{\gamma}\right) \left(\frac{\Delta}{T}\right)^{1/\gamma} ,
\end{equation}
namely a power-law decay for large $T$ with an exponent of $1/\gamma$.

Similarly to the Gumbel case with $\alpha>1$, the CS prediction is entirely off for large measurement times, starting from a power-law decay of a different exponent for $\gamma>2$, and ending with an exponential-like decay when $0<\gamma<2$. To compute this behavior, we rewrite the rescaled Fokker-Planck eigenvalue equation, based of Eq.~(\ref{equation: additional 5}),
\begin{equation}
	\frac{\dd^2}{\dd\chi^2} q_{\kappa}(\chi;\zeta) + \frac{\dd}{\dd\chi}\left[q_{\kappa}(\chi;\zeta) \frac{\gamma-1}{1-\chi} \right] + \kappa^2(z)q_{\kappa}(\chi;\zeta) = 0 ,
\end{equation}
with $\kappa=L\sqrt{\lambda}$ and $q_{\kappa}(\chi;\zeta)\dd\chi=p_k(x;z)\dd x$. This equation can be solved using Bessel functions similarly to the Fr\'echet example above, yielding the general solution
\begin{equation}
	q_{\kappa}(\chi;\zeta) = [\kappa(1-\chi)]^{\gamma/2}\left\{ A\text{J}_{\epsilon}[\kappa(1-\chi)] + B\text{J}_{-\epsilon}[\kappa(1-\chi)] \right\} , \quad \epsilon \equiv \frac{\gamma-2}{2} .
\end{equation}
The boundary conditions which we impose are reflection at $\chi=0$, which determines $B$ in terms of $A$, and absorption at $\chi=\zeta$, which yields $\kappa(\zeta)$, namely
\begin{equation}
	\left.\frac{\dd}{\dd\chi} q_{\kappa}(\chi;\zeta)\right|_{\chi=0} + (\gamma-1)q_{\kappa}(0;\zeta) = 0 , \quad q_{\kappa}(\zeta;\zeta) = 0 .
\end{equation}
These conditions give
\begin{equation}
	B = \frac{A\text{J}_{\epsilon+1}(\kappa)}{\text{J}_{-\epsilon-1}(\kappa)} ,
\end{equation}
and 
\begin{equation}
	\text{J}_{-\epsilon-1}(\kappa) \text{J}_{\epsilon}[\kappa(1-\zeta)] + \text{J}_{\epsilon+1}(\kappa) \text{J}_{-\epsilon}[\kappa(1-\zeta)] = 0 ,
\end{equation}
where the latter can be expanded for $\zeta\to1$,
\begin{equation}
\label{equation: weibull 3}
	\Gamma(\epsilon+1)\text{J}_{\epsilon+1}(\kappa) + \left[\frac{\kappa(1-\zeta)}{2}\right]^{\gamma-2} \Gamma(1-\epsilon)\text{J}_{-\epsilon-1}(\kappa) = 0 .
\end{equation}
Therefore, we see that one needs to separately consider two cases, as mentioned. For $\gamma>2$, the smallest root of Eq.~(\ref{equation: weibull 3}) is obtained for small $\kappa$, and thus by expanding we find
\begin{equation}
	\kappa_*^2(\zeta) \simeq \gamma(\gamma-2)(1-\zeta)^{\gamma-2} ,
\end{equation}
which decay to $0$ when $\zeta \to 1$. This gives for the CS limit of the DS EV CDF
\begin{equation}
	F_T^{\rm cs}(\zeta) \sim \exp\left[-\gamma(\gamma-2)(1-\zeta)^{\gamma-2}T_{\rm s} \vphantom{\frac{1}{1}}\right] \simeq \left[1-\gamma(\gamma-2)(1-\zeta)^{\gamma-2} \vphantom{\frac{1}{1}}\right]^{T_{\rm s}} ,
\end{equation}
where $T_{\rm s} \equiv DT/L^2$, and the subscript ``s" stands for ``scaled". This yields
\begin{equation}
\label{equation: weibull 4}
	1-\langle\zeta\rangle_{\rm cs} \simeq \Gamma\left(\frac{\gamma-1}{\gamma-2}\right) \left[\gamma (\gamma-2)T_{\rm s}\vphantom{\frac{1}{1}}\right]^{-1/(\gamma-2)} ,
\end{equation}
namely a different power-law decay with an exponent of $1/(\gamma-2)$. On the other hand, taking $0<\gamma<2$, we see that the second term of Eq.~(\ref{equation: weibull 3}) diverges when $\zeta\to 1$. Hence, the smallest root is found roughly as the first solution of $\text{J}_{-\epsilon-1}(\kappa)=0$. This is a positive number independent of $\zeta$, which means that the decay of $1-\langle\zeta\rangle_{\rm cs}$ is exponential-like for $0<\gamma<2$.

To illustrate these results, we first set $\gamma=2.5$ in Fig.~6. Plotting the deviation of the rescaled mean EV from its maximal possible value, $1-\langle\zeta\rangle$, versus the overall rescaled measurement time, $T_{\rm s}$, we see that even if one takes a small sampling time of $\Delta_{\rm s} \equiv D\Delta/L^2 = 10^{-3}$, the CS limit fails for large $T_{\rm s}$, and the IID RVs limit takes control of the EVs, with an underlying distribution which is the equilibrium measure. Note that we took $L=D=1$, such that $\{\zeta,\Delta_{\rm s},T_{\rm s}\}=\{z,\Delta,T\}$. In Fig.~6, the IID limit is given by Eq.~(\ref{equation: weibull 2}), the long-time asymptotics of the CS limit is given by Eq.~(\ref{equation: weibull 4}), and the exact CS limit is obtained by numerically solving the time-dependent Fokker-Planck equation, Eq.~(\ref{equation: additional 4}).

For the other regime, we set $\gamma=1$, giving the example of a particle freely diffusing in a box, as mentioned. Using Eq.~(\ref{equation: weibull 2}), the IID prediction for $1-\langle\zeta\rangle_{\rm iid}$ displays a power-law decay of $1/T_{\rm s}$ for large $T_{\rm s}$. Alternatively, the CS limit is obtained by solving the time-dependent Fokker-Planck equation, Eq.~(\ref{equation: additional 4}), with a left boundary condition of $\dd q(\chi,t_{\rm s}|\zeta)/\dd \chi|_{\chi=0}=0$, where $t_{\rm s} \equiv Dt/L^2$ is the rescaled time. This gives
\begin{equation}
	q(\chi,t_{\rm s}|\zeta) = \sum_{m=0}^{\infty} \frac{2}{\zeta} \cos\left[(1+2m) \frac{\pi\chi}{2\zeta}\right] \exp\left[-(1+2m)^2\frac{\pi^2t_{\rm s}}{4\zeta^2}\right] ,
\end{equation}
and so
\begin{equation}
\label{equation: weibull 5}
	F^{\rm cs}_{T_{\rm s}}(\zeta) = \int_0^{\zeta} \dd\chi\, q(\chi,T_{\rm s}|\zeta) = \sum_{m=0}^{\infty} \frac{4/\pi}{1+2m} \exp\left[-(1+2m)^2\frac{\pi^2T_{\rm s}}{4\zeta^2}\right] .
\end{equation}
Surprisingly, this representation for the EV CDF in the CS limit is incomplete, since in the truly continuous case, the (rescaled) upper bound on the particle's movement means that from a certain time and onward, the EV of any realization of the process becomes equal to $1$. This suggests that there is a $\delta(\zeta-1)$ contribution to the EV PDF with a time-dependent weight. This is further evidenced by noting that $F^{\rm cs}_{T_{\rm s}}(1)<1$ for $T_{\rm s}$ large enough, i.e. this distribution is not normalized. Incorporating this observation into the calculation is relatively straightforward. We simply write
\begin{equation}
	\frac{\dd}{\dd\zeta} F^{\rm cs,true}_{T_{\rm s}}(\zeta) = \frac{\dd}{\dd\zeta} F^{\rm cs}_{T_{\rm s}}(\zeta) + \left[1-F^{\rm cs}_{T_{\rm s}}(1)\right]\delta(\zeta-1) ,
\end{equation}
which is indeed normalized to unity. The deviation of the rescaled mean EV from its maximal possible value in the CS limit is then
\begin{equation}
\label{equation: weibull 6}
	1 - \langle\zeta\rangle_{\rm cs} = 1 - \int_0^1 \dd\zeta \left[ \frac{\dd}{\dd\zeta} F^{\rm cs,true}_{T_{\rm s}}(\zeta) \right] \zeta = \int_0^1 \dd\zeta\, F^{\rm cs}_{T_{\rm s}}(\zeta) .
\end{equation}
In the $T_{\rm s}\to\infty$ limit, we find
\begin{equation}
	1 - \langle\zeta\rangle_{\rm cs} \sim \frac{8}{\pi^3T_{\rm s}} \exp\left(-\frac{\pi^2}{4}T_{\rm s}\right) ,
\end{equation}
namely an exponential decay for large $T_{\rm s}$. Note that the first solution of $\text{J}_{-1/2}(\kappa_*)=0$ is exactly the effective $\kappa_*=\pi/2$ above.

Figure~7 shows the deviation of the rescaled mean EV from its maximal possible value, $1 - \langle\zeta\rangle$, versus the overall rescaled measurement time, $T_{\rm s}$, for a particle freely diffusing in a box. Note that we took $L=D=1$, such that $\{\zeta,\Delta_{\rm s},T_{\rm s}\}=\{z,\Delta,T\}$. In Fig.~7, the IID limit is given by Eq.~(\ref{equation: weibull 2}), and the CS limit is given by Eqs.~(\ref{equation: weibull 5}) and (\ref{equation: weibull 6}). It is clear that the CS limit fails for large $T_{\rm s}$, while the IID limit works excellently. This marks a qualitative difference between a DS with any finite $\Delta_{\rm s}$ to the CS limit of $\Delta_{\rm s}=0$, also for this example of particles in a box. Moreover, the required addition of the delta function suggests that the Weibull universality class is no longer an attractor for the continuous process, even though the IID limit does belong there. This change in the basin of attraction has interesting consequences, and is a worthy subject for a future research.


\begin{thebibliography}{9999}
\bibitem{Limit}
Note that taking $\Delta \to 0$ as discussed before and shown in Eq.~(\ref{eq05}) yields the CS limit for any finite $z$. Likewise, the $z\to\infty$ limit in Eq.~(\ref{eq08}) holds for any finite $\Delta$, where the relevant threshold is determined by the crossover scale of $z^2\Delta \sim O(1)$. However, these limits do not commute, and so the large-$z$ limit of Eq.~(\ref{eq05}) is different from the small-$\Delta$ limit of Eq.~(\ref{eq08}).
\end{thebibliography}
\begin{thebibliography}{9999}
\bibitem[SM3]{Hermite2}
One can prove Eq.~(\ref{equation: hermite expansion 2}) by differentiating it with respect to $\mu$, and using $\He_n(x)=x\He_{n-1}(x)-\He_{n-1}'(x)$ together with Eq.~(\ref{equation: hermite expansion 1}) to show that the resulted expression holds. Verifying that Eq.~(\ref{equation: hermite expansion 2}) is satisfied for $\mu=0$ concludes the proof.
\end{thebibliography}
\end{document}